\documentclass[aps,prd,a4paper,onecolumn,longbibliography,superscriptaddress,nofootinbib,floatfix,preprintnumbers,amsmath,amssymb,amsfonts,preprintnumbers]{revtex4-2}

\usepackage{bm}
\usepackage[usenames,dvipsnames]{color}
\usepackage{multirow}
\usepackage{graphicx}
\usepackage{adjustbox}
\usepackage{booktabs}
\usepackage{appendix}
\usepackage{url}
\usepackage{xspace}
\usepackage{tabularx}
\usepackage[normalem]{ulem} 
\usepackage{siunitx}
\usepackage[utf8]{inputenc}
\usepackage{xspace}
\usepackage[
        pdfstartview=XYZ,
        bookmarks=true,
        colorlinks=true,
        linkcolor=blue,
        urlcolor=blue, 
        citecolor=blue, % color of reference hyperlinks
        pdftex,
        bookmarks=true,
        breaklinks=true,
        linktocpage=true,    % page numbers as hyperlinks in toc
        hyperindex=true
]{hyperref}
\usepackage[capitalise]{cleveref}
\usepackage{orcidlink}
\usepackage{microtype}
\usepackage[inline]{enumitem}

\usepackage[a4paper]{geometry}      
\AtBeginDocument{
  \heavyrulewidth=.08em
  \lightrulewidth=.05em
  \cmidrulewidth=.03em
  \belowrulesep=.65ex
  \belowbottomsep=0pt
  \aboverulesep=.4ex
  \abovetopsep=0pt
  \cmidrulesep=\doublerulesep
  \cmidrulekern=.5em
  \defaultaddspace= .5em
  \setlength{\tabcolsep}{0.5em}
}

\newcommand{\MM}{\mathrm{MM}^2}

\begin{document}

%%%%%%%%%%%%%%%%%%%%%%%%%%%%%%%%%%%%%%%%%%%%%%%
\preprint{IFT-UAM/CSIC-23-111}
\preprint{IFIC/23-40, FTUV-23-0823.4331}
%%%%%%%%%%%%%%%%%%%%%%%%%%%%%%%%%%%%%%%%%%%%%%%

\vspace*{0.5cm}

%%%%%%%%%%%%%%%%%%%%%%%%%%%%%%%%%%%%%%%%%%%%%%%
% TITLE & DATE
%%%%%%%%%%%%%%%%%%%%%%%%%%%%%%%%%%%%%%%%%%%%%%%
\title{Discovering Long-lived Particles at DUNE 
}

%
% \date{\today}
%%%%%%%%%%%%%%%%%%%%%%%%%%%%%%%%%%%%%%%%%%%%%%%

%%%%%%%%%%%%%%%%%%%%%%%%%%%%%%%%%%%%%%%%%%%%%%%
% AUTHORS
%%%%%%%%%%%%%%%%%%%%%%%%%%%%%%%%%%%%%%%%%%%%%%%
\author{Pilar Coloma}
\email{pilar.coloma@ift.csic.es}
\affiliation{Instituto de F\'isica Te\'orica UAM-CSIC\\ Calle Nicol\'as Cabrera 13--15, Universidad Aut\'onoma de Madrid, 28049 Madrid, Spain} 

\author{Justo Mart\'in-Albo}
\email{justo.martin-albo@ific.uv.es}
\affiliation{Instituto de F\'isica Corpuscular (IFIC), CSIC \& Universitat de Val\`encia \\ Calle Catedr\'atico Jos\'e Beltr\'an, 2, 46980 Paterna, Valencia, Spain}

\author{Salvador Urrea}
\email{salvador.urrea@ific.uv.es}
\affiliation{Instituto de F\'isica Corpuscular (IFIC), CSIC \& Universitat de Val\`encia \\ Calle Catedr\'atico Jos\'e Beltr\'an, 2, 46980 Paterna, Valencia, Spain}

\begin{abstract}
Long-lived particles (LLPs) arise in many theories beyond the Standard Model. These may be copiously produced from meson decays (or through their mixing with the LLP) at neutrino facilities and leave a visible decay signal in nearby neutrino detectors. We compute the expected sensitivity of the DUNE liquid argon (LAr) and gaseous argon (GAr) near detectors (ND) to light LLP decays. In doing so, we determine the expected backgrounds for both detectors, which have been largely overlooked in the literature, taking into account their angular and energy resolution. We show that searches for LLP decays into muon pairs, or into three pions, would be extremely clean. Conversely, decays into two photons would be affected by large backgrounds from neutrino interactions for both near detectors; finally, the reduced signal efficiency for $e^+ e^-$ pairs leads to a reduced sensitivity for ND-LAr. Our results are first presented in a model-independent way, as a function of the mass of the new state and its lifetime. We also provide detailed calculations for several phenomenological models with axion-like particles (coupled to gluons, to electroweak bosons, or to quark currents). Some of our results may also be of interest for other neutrino facilities using a similar detector technology (e.g. MicroBooNE, SBND, ICARUS, or the T2K Near Detector). 
\end{abstract}

\maketitle

\newpage

\tableofcontents

\newpage

%%%%%%%%%%%%%%%%%%%%%%%%%%%%%%%%%%%%%%%%%%%%%%%%%%
\section{Introduction}
While the evidence pointing to the existence of physics \emph{beyond the Standard Model} (BSM) is overwhelming, the new physics has so far eluded its discovery in direct-detection experiments and colliders. Nevertheless, until very recently our experimental strategy was mostly focused on unveiling the existence of new states with masses at or above the electroweak scale. Interestingly, current constraints are successfully evaded by a plethora of BSM physics models containing light, feebly interacting states that offer viable solutions to most open problems in the Standard Model (SM) and avoid large corrections to the Higgs mass. Popular examples of this kind of models include those with heavy neutral leptons (HNL), which could explain the observed pattern of neutrino masses and mixing as well as the observed baryon asymmetry of the universe \cite{Akhmedov:1998qx, Asaka:2005pn, Drewes:2017zyw}; or models with a rich dark sector, which offer novel candidates for dark matter and may be connected to the SM through renormalizable portals at low energies (see, e.g., Refs.~\cite{Holdom:1985ag,Patt:2006fw, Batell:2009di,Falkowski:2009yz}). Thanks to their weak interactions with the SM, models of this sort typically include new particles that are long-lived and decay to SM states with a significant branching ratio.

The existence of light, feebly interacting, unstable states can be probed in multiple ways, from their impact on cosmological observables and astrophysical objects, to direct or indirect signals in laboratory experiments and colliders. In the case of unstable particles with masses in the $\mathcal{O}(0.1)$--$\mathcal{O}(10)$~GeV range, searches at fixed-target experiments typically offer the best constraints (see, for example, Refs.~\cite{Batell:2022dpx,Antel:2023hkf} for recent reviews). The key in this case is that, once produced, a \emph{long-lived particle} (LLP) may propagate over tens or even hundreds of meters before decaying into visible final states in nearby detectors. 

In recent years, the experimental search for LLPs has received considerable attention from the neutrino community. Accelerator-based neutrino experiments are entering a precision era with the primary goal of discovering CP violation in the lepton sector, but they also offer all the necessary ingredients to conduct sensitive fixed-target searches: namely, high-intensity proton beams producing large fluxes of mesons, as well as versatile near detectors. The current generation of accelerator-based neutrino oscillation experiments set already some of the leading constraints for certain LLP models (including dark scalars~\cite{Batell:2019nwo, MicroBooNE:2021usw, MicroBooNE:2022ctm}, axion-like particles~\cite{Coloma:2022hlv, Bertuzzo:2022fcm, ArgoNeuT:2022mrm}, or HNLs~\cite{Arguelles:2021dqn,Kelly:2021xbv,T2K:2019jwa,MicroBooNE:2022ctm,MicroBooNE:2019izn}, among others). These will surely be improved over the next decade by the two upcoming new-generation long-baseline neutrino oscillation experiments, Hyper-Kamiokande \cite{Hyper-Kamiokande:2018ofw} and the Deep Underground Neutrino Experiment (DUNE) \cite{DUNE:2020lwj}, currently under construction.

In this work, we highlight the unique opportunity offered in this regard by DUNE, which will be exposed to the LBNF neutrino beam~\cite{DUNE:2016evb}. In addition to its very high intensity, which leads to a considerably higher flux of pions and kaons with respect to other facilities, the high proton energy available at LBNF (120~GeV) will allow the production of a significant flux of heavier resonances such as $D$ mesons. This will enable the DUNE experiment to provide leading constraints on LLPs with masses above the kaon mass, a window that is otherwise challenging to explore for laboratory experiments.

In order to make our results as model-independent as possible, we will present them as a function of the mass of the new state and its lifetime, in line with our past works~\cite{Arguelles:2019ziu, Coloma:2019htx, Coloma:2022hlv, Coloma:2023adi} (see also Refs.~\cite{Batell:2023mdn, Contino:2020tix, Costa:2022pxv} for related works that also follow a model-independent approach). However, in order to put the expected sensitivities of DUNE in context and to ease their comparison to present limits, it is useful to consider a specific model. Thus, we will also consider models with pseudoscalar particles, which arise naturally as pseudo-Nambu-Goldstone bosons of a spontaneous global symmetry breaking, and are therefore ubiquitous in extensions of the SM. These particles are often referred to as \emph{axion-like particles} (ALPs), since the best-motivated example is the QCD axion~\cite{Weinberg:1977ma,Wilczek:1977pj}. Generic models with light unstable pseudoscalars may lead to a wide set of new physics signals in neutrino detectors depending on their couplings to the SM, including ALP decays into pair of electrons, muons, photons, or multiple mesons. The DUNE experiment will offer the possibility to study many of these, thanks to its highly-capable suite of near detectors (ND) \cite{DUNE:2021tad}, which will include both a liquid argon (LAr) and a gaseous argon (GAr) time projection chamber (TPC). In particular, while many of the constraints on ALPs rely on their coupling to photons and electrons, ALP couplings to muons are subject to fewer limits. For a wide class of models, these can be particularly relevant in the mass window above $2\thinspace m_\mu$ where the decay channel $a\to\mu\mu$ dominates. Modern facilities operating with TPC detectors offer excellent opportunities to provide leading constraints for these scenarios, as pointed out in Refs.~\cite{Coloma:2022hlv, Co:2022bqq}. Moreover, as the imaging capabilities of TPCs allow the possibility of studying complex final states with multiple particles (a key advantage with respect to other detector technologies used in neutrino experiments), in this work we will also consider ALP decays leading to multiple pions (for similar studies, see Refs.~\cite{Kelly:2020dda, Jerhot:2022chi}).

The multiple possibilities offered by DUNE to search for LLP decays have been pointed out in the literature before (e.g., in Refs.~\cite{Brdar:2020dpr,Berryman:2019dme,Brdar:2022vum,Capozzi:2021nmp,Bakhti:2018avv,Co:2022bqq,Dev:2021qjj, Ballett:2019bgd, Coloma:2020lgy,Batell:2023mdn,Kelly:2020dda,Breitbach:2021gvv,Dev:2021qjj,Krasnov:2019kdc,Jerhot:2022chi,Dev:2021qjj}). In most of these studies the effect of the background has been neglected, arguing that it could potentially be reduced to a negligible level by means of appropriate selection cuts. However, it should be kept in mind that the near detectors available at neutrino facilities are exposed to a high-intensity neutrino flux. The whole DUNE ND suite, for instance, will register more than 100 million neutrino interactions per year~\cite{DUNE:2021tad}. Even though the kinematic features of neutrino-nucleus scattering events are different than those expected from an LLP decay, reducing the expected background to a negligible level is a daunting task, as it was shown, for instance, in Refs.~\cite{Breitbach:2021gvv, Ballett:2019bgd} for the HNL scenario. The final states expected in the case of HNL decays through mixing are, however, different from those expected for other LLPs. In this work, we compute the expected backgrounds for generic final states involving two photons, two leptons, and multiple pions, with no missing energy. Thus, our background study is, a priori, applicable to BSM extensions including light vectors, scalars, or pseudoscalars. 

In principle, considering event rates from LLP-argon interactions at the ND, in addition to LLP decays, may provide additional sensitivity to certain BSM scenarios. This is specially so in regions of the parameter space corresponding to very long lifetimes, since the probability for an LLP to decay inside the detector becomes heavily suppressed. As pointed out in Refs.~\cite{Brdar:2020dpr,Brdar:2022vum,Capozzi:2023ffu}, the inclusion of scattering events for ALPs could be useful to close the so-called cosmological triangle; however, for ALPs above the MeV scale, the corresponding bounds are considerably worse than those obtained from decay searches (e.g., see Fig.~3 in Ref.~\cite{Brdar:2020dpr}). In our case, the study of scattering signals would demand a more involved simulation and analysis of the relevant backgrounds, falling outside the scope of the present work. 

This article is structured as follows. Section~\ref{Section:Production} describes the computation of the decay signal of a generic LLP. It also summarizes the main details of the  ALP benchmark models considered in this work, where we assume that the ALP is coupled to the SM with different sets of effective operators. Current constraints on these benchmark models are then summarized in Section~\ref{sec:bounds} (as well as in Appendix~\ref{app}). Next, we discuss our evaluation of the backgrounds in Section~\ref{sec:bg}, before presenting our results in Section~\ref{sec:results}. We summarize and conclude in Section~\ref{sec:conclusions}.

%%%%%%%%%%%%%%%%%%%%%%%%%%%%%%%%%%%%%%%%%%%%%%%%%%
\section{Production and decay of long-lived particles}
\label{Section:Production}
The computation of the number of decays of long-lived particles (LLP) in detectors located at a certain distance can be carried out as follows. We start from a Monte Carlo simulation of the production of parent mesons (technical details regarding these simulations are provided in Sec.~\ref{sec:bg}). The number of mesons can be written in terms of the number of \emph{protons on target} (PoT) and the meson yield per PoT ($Y_M$) as
\begin{equation}
\label{eq:mesons}
    \frac{d n_M}{dE_M d\Omega_M} = N_\mathrm{PoT} Y_M \frac{d^2 \rho_M}{dE_M d\Omega_M},
\end{equation}
where $d^2 \rho_M/(dE_M d\Omega_M)$ is the probability that a given meson is produced with energy in the interval $[E_M, E_M+dE_M]$ and with a trajectory defined by a solid angle in $[\Omega_M, \Omega_M+d\Omega_M]$. 
Using Eq.~(\ref{eq:mesons}), the expected number of particles $a$ produced in the decay $M\to a + \ldots$, with a branching ratio $\mathrm{BR}(M \rightarrow a)$, can be computed as
\begin{equation}
\label{eq:phi_aND}
    \frac{d n_a}{dE_a d\Omega_a} = 
    \mathrm{BR}(M \rightarrow a) \; N_\mathrm{PoT} Y_M
    \int d E_M \int d\Omega_M \frac{d^2 \rho_M}{d E_M d\Omega_M} \frac{d^2 \rho^{M\to a} (E_M, \Omega_M)}{dE_a d\Omega_a},
\end{equation}
where $d^2 \rho^{M\to a}/(dE_a d\Omega_a)$ stands for the differential probability that a meson will produce an LLP $a$ with energy and trajectory defined by $E_a$ and $\Omega_a$. As we will see, in certain scenarios the LLP may also be produced directly through their mixing with SM mesons. In this case, the LLP differential flux can be approximated by the meson flux in Eq.~\eqref{eq:mesons}, rescaled with the corresponding mixing angle accordingly (which depends on the masses of the two particles involved). 

At this point, it is important to note that the meson fluxes depend on the production point (inside the target and in the decay volume region). Similarly, the detector acceptance depends on the production point of the LLP, since this determines whether its trajectory crosses the detector. In what follows, in order to simplify our notation, we remove the dependence with the meson and the LLP production points; we stress, however, that in our numerical calculations this dependence has been fully accounted for. 

Once it has been produced, the LLP may live long enough to propagate to the detector and decay inside, with a probability that depends on its decay length boosted to the lab frame: $L_a = c\,\tau_a\,\gamma_a\,\beta_a$, where $\tau_a$ is the lifetime of the particle at rest, while $\beta_a$ and $\gamma_a$ are the boost factors. Such probability reads:
\begin{equation}\label{Eq:probability}
P_{\mathrm{decay}} (\Omega_a, E_a, c\tau_a / m_a) = 
e^{-\ell_\mathrm{det}/L_a} \cdot
\left( 1 - e^{-\Delta\ell_\mathrm{det}/L_a}\right) \, ,
\end{equation}
where $\ell_\mathrm{det}$ is the propagation distance before the particle enters the detector, and $\Delta \ell_\mathrm{det}$ is the length of the intersection between the trajectory of the particle and the detector (which in most cases approximately coincides with the detector length along the beam axis). Note that both quantities depend on the solid angle $\Omega_a$, as well as on the production point of the LLP. 

Eventually, the total number of LLP decays inside the detector into a given decay channel $ch$ is obtained after integration over the LLP variables, and multiplying by the corresponding branching ratio and the detector efficiency for that channel, $\epsilon_{ch}$:
\begin{equation}
    \label{eq:Ndec}
    N_{dec, ch} = \epsilon_{ch} \; \mathrm{BR}(a \to ch) \int dE_a \int_{\Omega_\mathrm{det}} d\Omega_a P_\mathrm{decay}(\Omega_a, E_a, c\tau_a / m_a) \frac{dn_a}{dE_a d\Omega_a} \, ,
\end{equation}
where the produced LLP flux and the decay probability are given in Eqs.~(\ref{eq:phi_aND}) and (\ref{Eq:probability}), respectively, and the integral in solid angle is performed taking into account only those trajectories within the angular acceptance of the detector, $\Omega_\mathrm{det}$.

\begin{figure}
\centering
\includegraphics[width=0.495\textwidth]{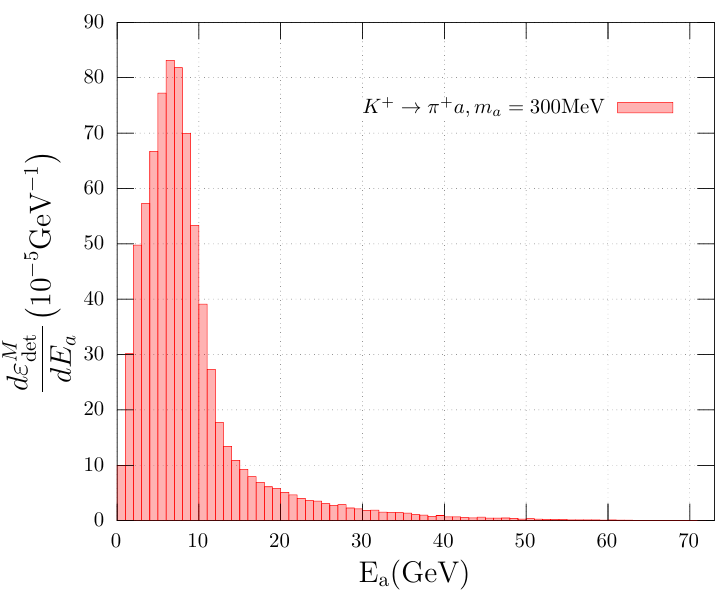}
\includegraphics[width=0.495\textwidth]{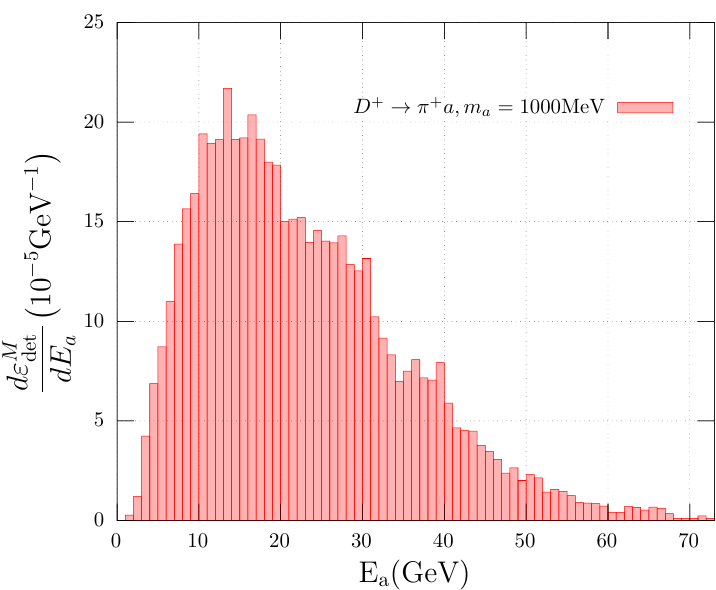}
\caption{\label{fig:Acceptance} Detector acceptance as a function of the energy of the ALPs. On the left panel, the ALPs have a mass of 300~MeV and originate from the decay $K^+ \to \pi^+ a$. On the right panel, the ALPs have a mass of 1~GeV and are produced via $D^+ \to \pi^+ a$.}
\end{figure}

Even though Eq.~(\ref{eq:Ndec}) is exact (and it is, indeed, what we use in our computation of the number of signal events), it might not be very illuminating. We can derive a simpler expression noting that for the experimental setup considered here: 
\begin{enumerate*}[label=(\roman*)]
    \item the distance to the detector is much longer than the size of the target (where most of the mesons are produced);
    \item the detector size is much smaller than the distance traveled by the LLP before reaching the detector;
    \item the angular acceptance of the detector is small, which mostly selects particles traveling along the beam axis.
\end{enumerate*}
These allow us to assume that all LLPs are approximately produced at the same point and to neglect the dependence of $\ell_\mathrm{det}$ and $\Delta \ell_\mathrm{det}$ with the trajectory of the LLP. Under these approximations, we obtain:
\begin{equation}
\label{eq:Ndec-approx}
N_{dec, ch} \simeq \epsilon_{ch} \; \mathrm{BR}(a \to ch) \mathrm{BR}(M \to a) N_{PoT} Y_M \int dE_a P_\mathrm{decay}(E_a, c\tau_a / m_a) \frac{d\varepsilon^M_\mathrm{det} (m_a)}{dE_a} \, ,
\end{equation}
where $d\varepsilon^M_{det}/dE_a$ is the angular acceptance of the detector for an LLP with energy $E_a$. Since the detector acceptance relies on the boost of the LLP in the lab frame, it will depend on the energy and momentum distribution of the parent meson $M$, on the mass of the LLP, and on whether the LLP is produced in a two-body or a three-body decay. As illustration, the differential detector acceptance is shown in Fig.~\ref{fig:Acceptance} as a function of the energy of the LLP, for $K$ and $D$ two-body decays and for two representative values of the mass of the LLP. As can be seen from this figure, the accepted flux ranges from a few GeV to tens of GeV, with some dependence on the parent meson and the LLP mass.

%%%%%%%%%%%%%%%%%%%%%%%%%%%%%%%%%%%%%%%%%%%%%%%%%%
%%%%%%%%%%%%%%%%%%%%%%%%%%%%%%%%%%%%%%%%%%%%%%%%%%

Up to this point, the discussion is model-independent and applies to any LLP produced from meson decays at a neutrino beamline. The final sensitivity will be fully determined by the mass of the particle and its lifetime, and it will be optimal for values of $c\tau_a$ which maximize the decay probability in Eq.~(\ref{Eq:probability}). In order to compute the expected number of events, one just needs to know the parent meson and whether the LLP is produced through a two-body or a three-body decay. Nevertheless, in a given model the production branching ratio and the decay width will typically be partially correlated and depend on a set of common parameters, which will also enter the decay branching ratio into a given channel. The key requirement for these searches to succeed is that the lifetime of the LLP should be long enough so that it reaches the detector before decaying. This pushes into the weakly interacting limit, which demands small couplings and, therefore, typically small production branching ratios. Consequently, in order to reach the maximum sensitivity region, a trade-off must be found in terms of the production branching ratios and lifetime, which is not always easy to do and depends severely on the model. The second aspect to consider is that specific benchmarks will translate into preferred decay channels, which ultimately determine the experimental strategy to follow in order to maximize the signal-to-background ratio. 

In the rest of this section, we will focus on three \emph{benchmark models} in the context of ALPs. This will allow us to write specific expressions, in terms of the model parameters, for the ALP production and decay branching ratios, as well as its lifetime. As we will see, the three models considered would lead to very distinct phenomenology in an experimental search.

%%%%%%%%%%%%%%%%%%%%%%%%%%%%%%%%%%%%%%%%%%%%%%%%%%
\subsection{Benchmark Model I: Gluon Dominance scenario}
We start by considering an anomalous coupling between ALPs and the gluon field, leading to a higher-dimensional effective operator. This has been referred to in the literature as the \emph{gluon dominance} scenario~\cite{Beacham:2019nyx}, and is one of the main targets for the vast suite of experiments searching for long-lived particles (see, e.g., Refs.~\cite{Agrawal:2021dbo,Antel:2023hkf,Goudzovski:2022vbt}). The sensitivity of DUNE to ALPs in this scenario has been studied previously in Refs.~\cite{Kelly:2020dda,Jerhot:2022chi} assuming that backgrounds could be reduced to a negligible level for the most relevant decay channels. Here, we revisit their limits taking into consideration the expected background levels, and slight improvements (outlined below) on the calculation of the ALP decay widths. Following the same normalization convention as in Ref.~\cite{Beacham:2019nyx}, the relevant ALP interaction Lagrangian reads:
\begin{equation}
\label{eq:GGdual}
    \delta\mathcal{L}_{a, int} = c_G \mathcal{O}_G = \frac{\alpha_s}{8\pi f_a}a G^b_{\mu\nu} \widetilde{G}^{b\mu\nu} 
    %= \frac{g^2_s}{8 f_G} a G^b_{\mu\nu} \widetilde{G}^{b\mu\nu}  
    \, , 
\end{equation}
where $G^b_{\mu\nu}$ is the gluon field strength, $\widetilde{G}^{b\mu\nu} \equiv \frac{1}{2}\epsilon^{\mu\nu\rho\sigma}G_{b\rho\sigma}$, with $\epsilon^{0123} = 1$. Also, $\alpha_s \equiv g^2_s /(4\pi)$, and $g_s$ stands for the strong coupling constant. 

\begin{figure}
\centering
\includegraphics[width=\textwidth, trim=0 20 0 0]{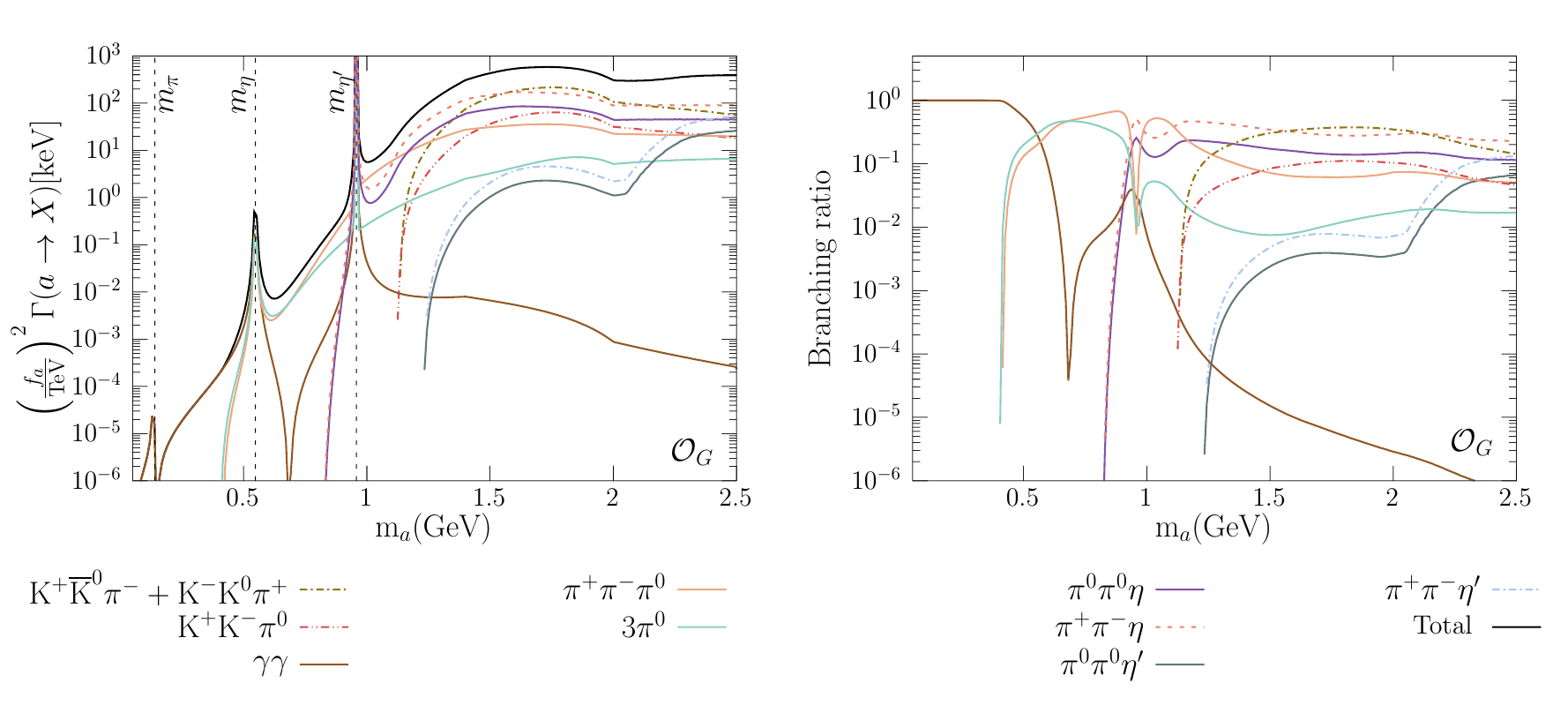}
\caption{\label{fig:BR-cgg} Main ALP decay widths (left) and branching ratios (right) for the gluon dominance scenario in Eq.~\eqref{eq:GGdual}, as a function of its mass. These have been computed following Ref.~\cite{Cheng:2021kjg} (see also Ref.~\cite{Aloni:2018vki}). }
\end{figure}

In this scenario, the ALPs would be mostly produced directly through their mixing with neutral pseudoscalar mesons ($\pi^0$, $\eta$, $\eta^\prime$) below 1~GeV, and from gluon fusion for masses above this value. Here, we use the calculation from Ref.~\cite{Kelly:2020dda}. As for their decays, for $m_a < 3m_\pi$ the only available decay channel is $a\to\gamma\gamma$, since CP conservation forbids the decay into two pions, whereas for higher masses, hadronic decay modes (with three or more mesons in the final state) will dominate. ALP interactions with pseudoscalar mesons may be described using chiral perturbation theory including ALPs (a$\chi$PT), but a proper treatment of the interactions between multiple mesons requires the inclusion of Vector Meson Dominance (VMD) terms in the Lagrangian~\cite{Fujiwara:1984mp}. Throughout this work, we follow Refs.~\cite{Aloni:2018vki, Cheng:2021kjg}, where a data-driven method (similar to the one developed in Ref.~\cite{Ilten:2018crw}) is used to successfully describe ALP interactions within this framework up to relatively high ALP masses ($m_a \lesssim 3~\mathrm{GeV}$). The effective $a\chi$PT Lagrangian is then matched onto the perturbative QCD (pQCD) Lagrangian. Figure~\ref{fig:BR-cgg} shows the decay widths and branching ratios for the most relevant decay channels in this scenario, computed following Ref.~\cite{Cheng:2021kjg} (see also Ref.~\cite{Aloni:2018vki}). The sharp features observed in the decay width stem from the large mixing between the ALP and the neutral pseudoscalar mesons whenever $m_a \sim m_{\pi^0}, m_\eta, m_{\eta^\prime}$, indicated by the dotted vertical lines.  

%%%%%%%%%%%%%%%%%%%%%%%%%%%%%%%%%%%%%%%%%%%%%%%%%%
%%%%%%%%%%%%%%%%%%%%%%%%%%%%%%%%%%%%%%%%%%%%%%%%%%

%%%%%%%%%%%%%%%%%%%%%%%%%%%%%%%%%%%%%%%%%%%%%%%%%%
\subsection{Benchmark Model II: ALPs coupled through electroweak operators }
Next, we consider an effective Lagrangian that includes ALP couplings to electroweak operators through effective operators arising at a high-energy scale $\Lambda$ (which we set to $\Lambda = f_a =1~\mathrm{TeV}$):
\begin{align}
\delta\mathcal{L}_{a, int} = c_\phi\mathcal{O}_\phi + c_B \mathcal{O}_B + c_W \mathcal{O}_W = c_{\phi} \frac{\partial^\mu a }{f_a}\phi^\dagger i\overleftrightarrow{D}_\mu \phi  
- c_{B} \frac{a}{f_a}B_{\mu\nu}\widetilde{B}^{\mu\nu} 
- c_W\frac{a}{f_a}W^I_{\mu\nu}\widetilde{W}_I^{\mu\nu} \, ,
\end{align}
where $B$ and $W^I$ stand for the EW vector bosons, $\phi$ is the Higgs doublet and $\phi^\dagger i\overleftrightarrow{D}_\mu \phi \equiv \phi^\dagger \left( iD_\mu\phi \right) - \left( i D_\mu \phi\right)^\dagger \phi $. It is worth mentioning that $\mathcal{O}_\phi$ can be traded by a set of (flavor-conserving) fermionic operators by means of a hypercharge rotation, as shown in Ref.~\cite{Georgi:1986df}. This finally leads to:
\begin{equation}
\label{eq:Lag-EW}
\delta\mathcal{L}_{a, int} = \frac{\partial_\mu a}{2f_a} \sum_f c_{ff} \bar f \gamma^\mu \gamma_5 f 
- c_{B} \frac{a}{f_a}B_{\mu\nu}\widetilde{B}^{\mu\nu} 
- c_W\frac{a}{f_a}W^I_{\mu\nu}\widetilde{W}_I^{\mu\nu} \, ,
\end{equation}
where sum over $f$ extends to quarks and charged leptons, with $c_{ff}=c_{\phi}$ for down-type quarks and charged leptons, and $c_{ff} = -c_\phi$ for up-type quarks. 

Although the operators included in Eq.~(\ref{eq:Lag-EW}) are flavor conserving, the $\mathcal{O}_{\phi}$ and $\mathcal{O}_W$ operators can induce flavor-changing neutral current (FCNC) processes at one loop, as detailed, for instance, in Refs.~\cite{Izaguirre:2016dfi,Gavela:2019wzg, Guerrera:2021yss}. For this scenario, the mechanism of production we consider will be kaon decays to ALPs ($K \to \pi a$), in line with our past work in Ref.~\cite{Coloma:2022hlv}. The corresponding width can be computed using the a$\chi$PT Lagrangian~\cite{Bauer:2021wjo,Georgi:1986df}: 
\begin{equation}
\label{eq:BKpia}
\Gamma(K\to\pi a) = \frac{m_K^3 |[k_Q(\mu)]_{sd}|^2}{64\pi f_a^2}
f_0\left(m_a^2\right)\lambda^{1/2}(1, m_a^2/m_K^2, m_\pi^2/m_K^2) \left(1-\dfrac{m_{\pi}^2}{m_{K}^2}\right)^2\,,
\end{equation}
where $f_0$ is the scalar form factor\footnote{For the range of masses we are considering, $f_0\left(m_a^2\right)$ can be closely approximated to 1, see Ref.~\cite{Carrasco:2016kpy}.} and
\begin{eqnarray}
\label{eq:lambda}
  \lambda(a,b,c)  & = a^2 + b^2 + c^2 - 2ab - 2ac - 2bc \, .
\end{eqnarray}
In Ref.~\cite{Coloma:2022hlv} we computed the running of the couplings from $\Lambda = 1~\rm{TeV}$ down to $\mu = 2~\mathrm{GeV}$, following Ref.~\cite{Bauer:2020jbp}. This leads to the following matching condition for the effective coupling entering the decay width: 
\begin{equation}
\label{eq:keff-value}
\frac{[k_Q(2~\mathrm{GeV})]_{sd}}{V_{td}^*V_{ts}}\bigg|_{\Lambda=\rm{1 TeV}} \simeq  -9.7\times 10^{-3}c_W(\Lambda)  + 8.2\times 10^{-3}c_{\phi}(\Lambda)  - 3.5\times 10^{-5}c_{B}(\Lambda) \, .
\end{equation}
We note that a similar effective coupling is generated in this class of models for the decay $B \to K a$; however, the production of $B$ mesons is insufficient at DUNE and will not be considered here.

ALPs produced from kaon decays will have a kinematical threshold at $m_a < m_K - m_\pi \sim 355~\mathrm{MeV}$. In this mass window, the ALP can only decay into photons or light charged leptons pairs ($e$ and $\mu$).  The decay width for the di-photon decay channel is given by
 \begin{align}
\label{eq:GammaPhoton}
\Gamma (a\to \gamma\gamma)& =  |c_{\gamma\gamma}|^2 \dfrac{m_a^3 }{4\pi f_a^2}\,,
\end{align}
where the effective coupling at low scales is given at one loop by~\cite{Gavela:2019wzg,Bauer:2017ris}
\begin{align}
\begin{split}
c_{\gamma \gamma } =  &c_W\,\Big[s_w^2\,+\frac{2\,\alpha}{\pi} B_2(\tau_W)\Big] +c_B\,c_w^2 
- c_{\phi} \,\frac{\alpha}{4\pi}\,\bigg( B_0 
- \frac{m_a^2}{m_{\pi}^2-m_a^2}\bigg)\, .
\end{split}
\label{eq:cgg}
\end{align}
Here, we have written $c_i \equiv c_i(\Lambda)$, $B_0$ and $B_2$ are loop functions (which can be found, for example, in Appendix B of Ref.~\cite{Coloma:2022hlv}), $\tau_W = 4m_W^2/m_a^2$ and $\alpha$ is the fine-structure constant. 

Finally, the decay width into dilepton pairs is given by
\begin{equation}
\Gamma (a\to \ell^+\ell^-) =  
|c_{\ell\ell}|^2\dfrac{ m_a m_\ell^2 }{8\pi f_a^2} \sqrt{1-\dfrac{4 m_\ell^2}{m_a^2}} \, ,
\label{eq:GammaLepton}
\end{equation}
where  $c_{\ell\ell}$ has been computed at one loop, and at low energies ($\mu \sim 2~\mathrm{GeV}$) reads~\cite{Gavela:2019wzg,Bauer:2017ris}
\begin{equation}
c_{\ell\ell} = c_{\phi}+\frac{3\,\alpha}{4\pi} \left(\frac{3\,c_W }{s_w^2}+ \frac{5\,c_B}{c_w^2 } \right) \log \dfrac{f_a}{m_W} 
+\dfrac{6\, \alpha}{\pi}\left(c_B \, c_w^2 + c_W\, s_w^2 \right) \log \dfrac{m_W}{m_\ell} \,.
\label{eq:cll}
\end{equation}

\begin{figure}
\centering
\includegraphics[width=\textwidth, trim=0 20 0 0]{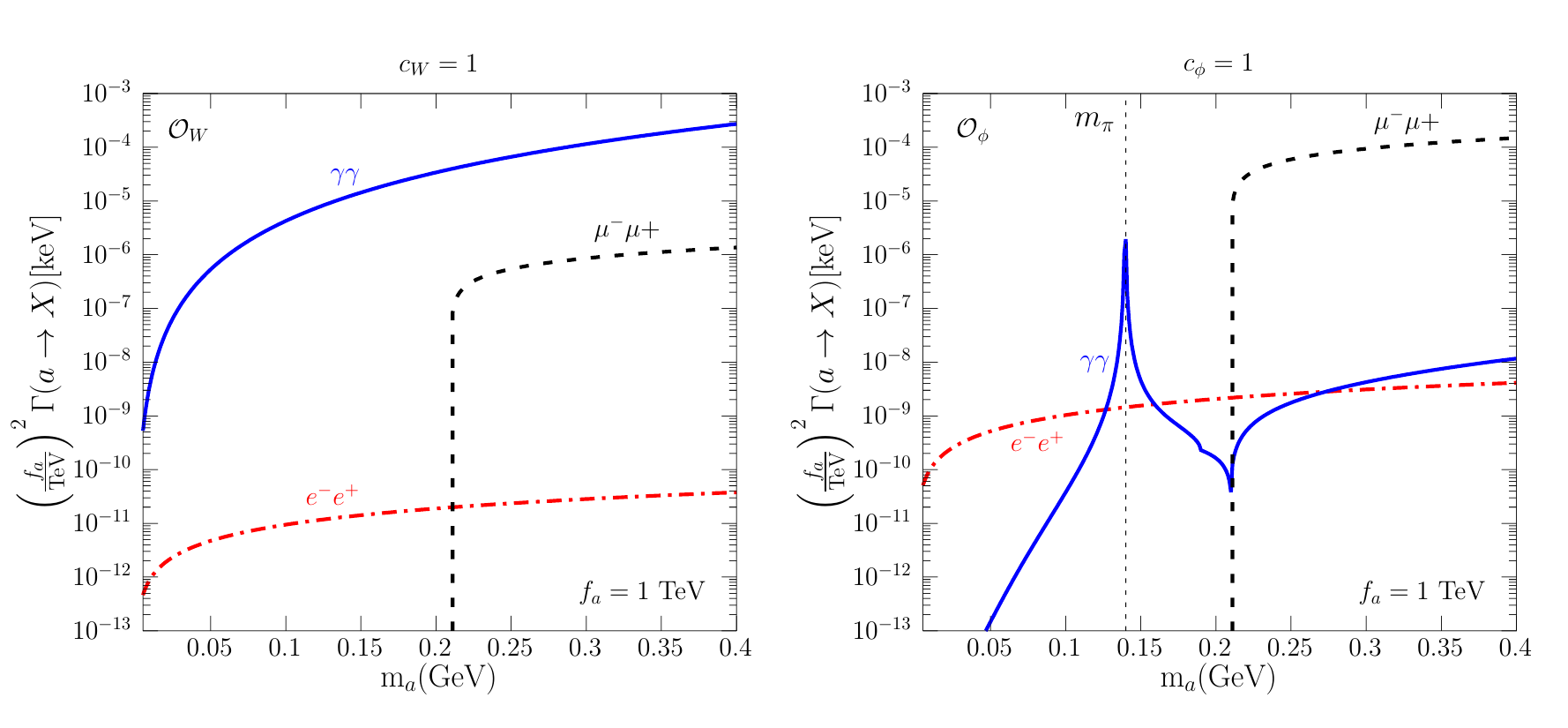}
\caption{\label{fig:BR-EW} Main decay widths for the ALP considered in our benchmark model II (BM-II), as a function of its mass $m_a$. Results for an ALP coupled predominantly through $\mathcal{O}_W$ are shown in the left panel, while the right panel shows similar results for $\mathcal{O}_\phi$. }
\end{figure}

By direct observation of Eqs.~(\ref{eq:keff-value})--(\ref{eq:cll}), we note that:
\begin{itemize}
    \item $\Gamma(a\to \ell^+\ell^-)$ is suppressed by a factor of $m_{\ell}^2/m_a^2$ with respect to $\Gamma(a\to\gamma\gamma)$. Therefore, for similar values of $c_{\gamma\gamma}$ and $c_{\ell\ell}$, we expect $\Gamma(a\to\gamma\gamma) \gg \Gamma(a\to \ell^+ \ell^-)$.
    \item For ALPs masses $m_a > 2m_{\mu}$, the decay channel $a\to\mu^+\mu^-$ will completely dominate over the decay channel $a\to e^+ e^-$, since $\Gamma (a\to \ell^+\ell^-)\propto m_{\ell}^2$.
    \item Due to the suppression with $\alpha$, $\Gamma(a\to \ell^+\ell^-)$ is mostly induced by $\mathcal{O}_\phi$, while $\Gamma(a\to\gamma\gamma)$ is mostly induced by $\mathcal{O}_B c_w^2+ \mathcal{O}_W s_w^2$.
    \item As can be seen in Eq.~(\ref{eq:keff-value}), $\mathcal{O}_B$ has a subdominant effect in the ALP production. However, it can have a significant effect in the ALP decay width, affecting its lifetime. For simplicity, hereafter we set the coefficient of this operator to zero; however, we refer the interested reader to Refs.~\cite{Gavela:2019wzg, Coloma:2022hlv} for a related discussion on this issue.
\end{itemize}  
The corresponding decay widths are shown in Fig.~\ref{fig:BR-EW} for an ALP coupled predominantly through $\mathcal{O}_W$ (left panel) and $\mathcal{O}_\phi$ (right panel).

%%%%%%%%%%%%%%%%%%%%%%%%%%%%%%%%%%%%%%%%%%%%%%%%%%
%%%%%%%%%%%%%%%%%%%%%%%%%%%%%%%%%%%%%%%%%%%%%%%%%%

Finally, we conclude this subsection stressing that the production of ALPs from decays of heavier mesons (e.g., $D^{+} \rightarrow \pi^{+} a$) using couplings to electroweak operators would require to generate the effective flavor-violating coupling $[k_Q]_{cu}$ via loop. However, proceeding analogously as for the kaon case, it is easy to see that the corresponding coupling for $D$ decays will be severely suppressed, since
\begin{equation}
\frac{[k_Q]_{cu}}{[k_Q]_{sd}} \propto \frac{| V^*_{ub} V_{cb}| }{| V^*_{td} V_{ts} | } \frac{y_b^2}{y_t^2} \sim 2.5 \times 10^{-4}\, \quad (\mathrm{for}~\mathcal{O}_W, \mathcal{O}_B, \mathcal{O}_\phi). \nonumber
\end{equation} 
Therefore, no sensitivity to these operators is expected at DUNE for $m_a > m_K$. 

\subsection{Benchmark Model III: Charming ALPs}
\label{sec:charmALPs}

Here, we consider a different set of operators that could induce ALP production from $D$ decays, which would allow to probe even heavier ALPs. This is the case, for example, of effective operators coupling the ALP to a quark current with off-diagonal couplings in flavor space (see, e.g., the discussion in Ref.~\cite{MartinCamalich:2020dfe} and references therein). The inclusion of couplings to quarks implies, however, that the ALP will have a significant decay width into hadronic final states. Consequently, its lifetime and branching ratios will show a significant dependence on the assumed Wilson coefficients and it becomes necessary to choose a particular texture in flavor space. As an illustrative example, we will consider ALPs that couple to right-handed up-quark currents as in Ref.~\cite{Carmona:2021seb}, dubbed as \emph{charming ALPs}:
\begin{equation}
\label{eq:LagCharming}
\delta\mathcal{L}_{a, int} = c_{u_R} \mathcal{O}_{u_R} = \sum_{i,j} \frac{\partial_\mu a}{f_a}\left(c_{u_R}\right)_{i j} \bar{u}_{R i} \gamma^\mu u_{R j} \, ,
\end{equation}
where $c_{u_R}$ is a Hermitian matrix in flavor space, and $i,j=1,2,3$ are quark family indices. The flavor-violating coupling $[c_{u_R}]_{12}$ in Eq.~\eqref{eq:LagCharming} induces $D^{+} \rightarrow \pi^{+} a$, with a decay width given  by~\cite{Carmona:2021seb,Bauer:2021mvw} 
\begin{equation}
\label{eq:decayDalps}
\Gamma(D^+ \rightarrow \pi^+ a)=\frac{m_D^3\left|\left[c_{u_R}\right]_{12}\right|^2}{64 \pi f_a^2}\left[f_0^{D \pi}\left(m_a^2\right)\right]^2 
\lambda^{1 / 2}\left(1, m_a^2 / m_D^2, m_\pi^2 / m_D^2\right)\left(1-\frac{m_\pi^2}{m_D^2}\right)^2 ,
\end{equation}
where $\lambda$ is defined in Eq.~\eqref{eq:lambda} and $f_0^{D \pi}\left(m_a^2\right)$ is the scalar form factor evaluated at a momentum transfer $q^2 = m_a^2$, which we take from the lattice computation in Ref.~\cite{Lubicz:2017syv}. 

For ALPs produced in $D$-meson decays, we can probe a wide mass window up to $m_a = m_D - m_\pi \lesssim 1.7~\mathrm{GeV}$. As in the gluon dominance scenario, in this case the dominant decay modes will involve multiple mesons in the final state, which requires the use of the $a\chi$PT for masses below approximately 2.5-3~GeV. We note that the calculation in Ref.~\cite{Cheng:2021kjg} describes ALP interactions with mesons induced by purely diagonal couplings in flavor space. However, we can use it for the purposes of this work since the only off-diagonal couplings considered involve $c$ and $t$ quarks, which are irrelevant for decays in this mass region. In particular, we stress that as long as the ALP couples to up-quarks only, its decay widths are mostly determined by the $[c_{u_R}]_{11}$ coupling in Eq.~(\ref{eq:LagCharming}), while the remaining couplings only enter at subleading order for some channels (such as $a\to \gamma\gamma$).

\begin{figure}
\centering
\includegraphics[width=\textwidth, trim=0 30 0 0]{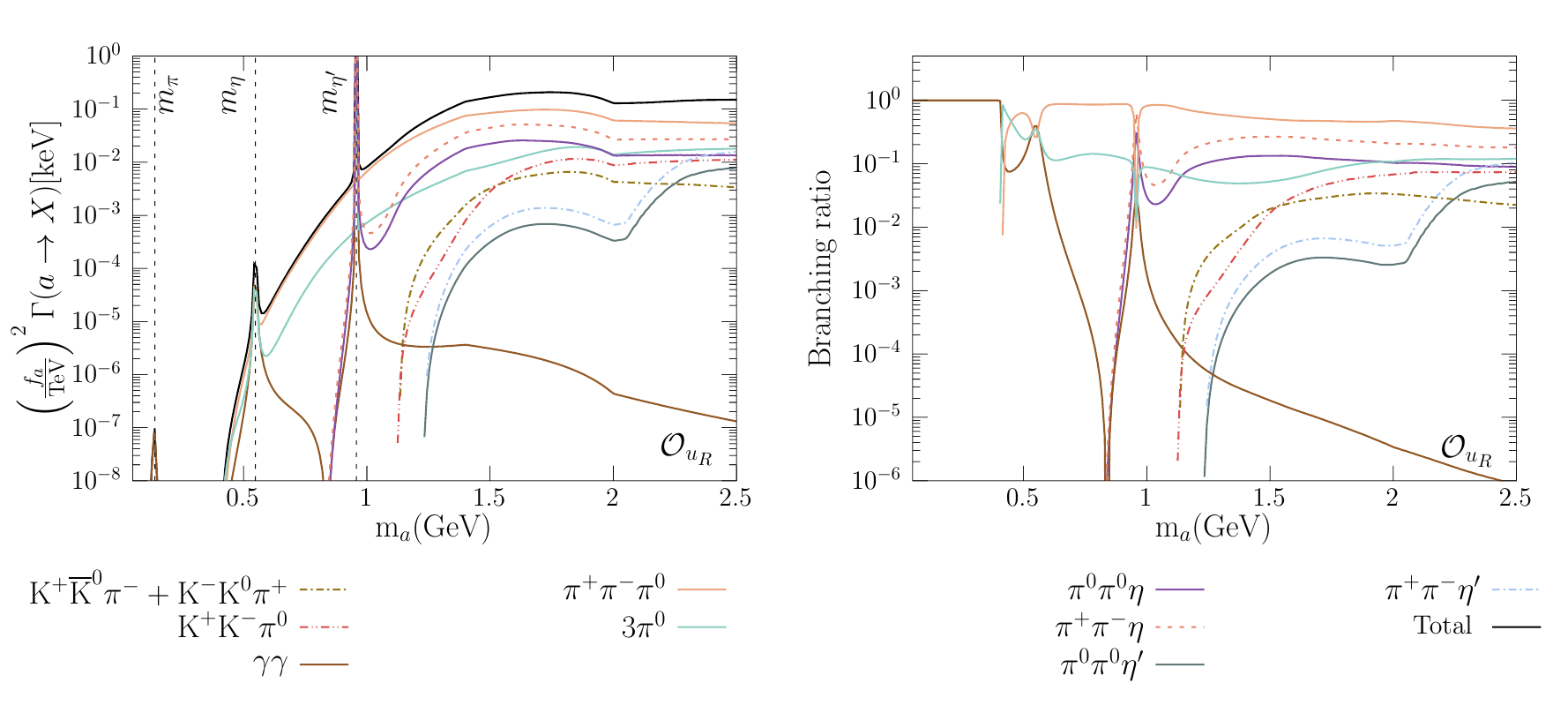}
\caption{\label{fig:charming_alps_decays} Main decay widths (left) and branching ratios (right) for the Charming ALP scenario, as a function of its mass. These have been computed following Ref.~\cite{Cheng:2021kjg}, for the couplings in Eq.~\eqref{eq:textureFN}.  
}
\end{figure}

In summary, for the purposes of this work, the phenomenology of this class of models is fully determined by $[c_{u_R}]_{12}$ (which controls the production in $D$ decays) and by $[c_{u_R}]_{11}$ (which controls the decay of the ALP). Hence, it becomes interesting to consider the expected size of these two couplings for well-motivated UV completions leading to the set of operators in Eq.~(\ref{eq:LagCharming}). The authors of Ref.~\cite{Carmona:2021seb} studied several possible UV completions of this scenario. Here, we highlight two of them, where the ALP emerges as a pseudo-Nambu-Goldstone boson (pNGB) of a spontaneously-broken symmetry:
\begin{itemize}
    \item The first one is a Froggatt-Nielsen (FN) model, where the ALP corresponds to what is usually called a \emph{flavon} or a \emph{familion}~\cite{REISS1982217,Feng:1997tn,Berezhiani:1990wn,Berezhiani:1990jj,Davidson:1981zd}.
    The model includes a complex scalar field $S$, whose radial component is identified with the ALP, and a new $U(1)$ flavor symmetry. The inclusion of higher-dimensional operators at a high-energy scale $\Lambda > f_a$ generates the operators in Eq.~(\ref{eq:LagCharming}) once the scalar acquires a vacuum expectation value, $\langle S \rangle = f_a$. As shown in Ref.~\cite{Carmona:2021seb}, an appropriate choice of the charges under the new symmetry is able to generate the correct up-quark masses (thus alleviating the \emph{flavor} problem) and would lead to
        \begin{equation}\label{eq:textureFN}
        c_{u_R}^\mathrm{FN} = \left(\begin{array}{ccc}
        2 & 3 \epsilon & 3 \epsilon^2 \\
        3 \epsilon & 1 & \epsilon \\
        3 \epsilon^2 & \epsilon & \epsilon^2
        \end{array}\right) \; ,
        \end{equation}
    where off-diagonal entries are controlled by $\epsilon=f_a/\Lambda \sim m_c/m_t$. 
    \item The second possibility is a dark-QCD (dQCD) model with a confining dark sector that contains $n_d$ dark flavors transforming under $SU(n_d)$. A dark confining sector would offer plausible candidates for dark matter among the dark hadrons and therefore is also well-motivated from the theoretical point of view. In addition, the ALP may be identified with one of the CP-odd states in the dark meson spectrum (a \emph{dark pion}, see e.g. Ref.~\cite{Cheng:2021kjg}). The operators in Eq.~(\ref{eq:LagCharming}) may be generated, for example, through a Yukawa interaction between SM quarks and dark quarks with a heavy scalar field, which is integrated out of the theory at low energies. Adopting the same assumptions and parameter values as in Ref.~\cite{Carmona:2021seb}, this leads to a texture for $c_{u_R}$ in the $u-c$ sector that is very similar to the one in Eq.~\eqref{eq:textureFN}, up to an overall rescaling factor. 
\end{itemize}
In summary, both of these UV completions would lead to $[c_{u_R}]_{12} / [c_{u_R}]_{11} \sim \mathcal{O}(0.01)$--$\mathcal{O}(0.03)$, with small differences depending on the particular scenario being considered. Fixing a texture is convenient since it allows to compute the corresponding ALP decay widths for all relevant decay channels, and sets the correlation between the ALP production rates and its lifetime. The corresponding branching ratios for the texture in Eq.~\eqref{eq:textureFN} are shown in Fig.~\ref{fig:charming_alps_decays}, where we have set the matching scale between the $a\chi$PT and the pQCD Lagrangians at 2.9~GeV following Ref.~\cite{Cheng:2021kjg}. We see that for $m_a < 3m_\pi$ the ALP decays exclusively to $a\rightarrow \gamma\gamma$. Conversely for $m_a > 3m_\pi$ the decay $a\rightarrow \pi^{+} \pi^{-} \pi^0$ rapidly dominates, as shown in Fig.~\ref{fig:charming_alps_decays}, with the exception of the regions neighbouring the mass values of $m_\eta$ and $m_\eta^{\prime}$. The resulting branching ratios for a dQCD-inspired model would be very similar to the ones shown here for the FN-inspired model, although the corresponding ALP lifetime would be longer. This would affect the statistics expected at the detector (inducing a change in sensitivity), but the results would otherwise be qualitatively very similar. Therefore in what follows for concreteness we will adopt the texture in Eq.~\eqref{eq:textureFN} to show our results for this benchmark scenario.  

%%%%%%%%%%%%%%%%%%%%%%%%%%%%%%%%%%%%%%%%%%%%%%%%%%
%%%%%%%%%%%%%%%%%%%%%%%%%%%%%%%%%%%%%%%%%%%%%%%%%%

We finalize this section by stressing that, while we have based our discussion on theoretically well-motivated examples, our results for this benchmark scenario will apply to a wider class of models, as long as the ALP does not couple to down quarks directly\footnote{Including a coupling to down quarks would mainly affect the calculation of the decay widths and branching ratios \cite{Cheng:2021kjg}.} and both $[c_{u_R}]_{11},[c_{u_R}]_{12}$ are generated.

%%%%%%%%%%%%%%%%%%%%%%%%%%%%%%%%%%%%%%%
\section{Previous constraints}
\label{sec:bounds}
%%%%%%%%%%%%%%%%%%%%%%%%%%%%%%%%%%%%%%%

This section contains a brief summary of the most relevant constraints for the two benchmark models under consideration. Most of the bounds from current and past experiments have been previously computed in the literature, and are just recasted here for the scenarios of interest. In the case of an ALP coupled through electroweak effective operators, a comprehensive discussion on the applicable bounds can be found e.g., in Refs.~\cite{Brivio:2017ije,Bauer:2021mvw,Goudzovski:2022vbt} (see also Ref.~\cite{Coloma:2022hlv}). For an ALP coupled to right-handed up-quarks, we follow the discussion in Ref.~\cite{Carmona:2021seb}, where the authors derived the most relevant bounds on this scenario. However in this case we have rederived some of the constraints here, finding significant differences as explained in more detail below. 

\subsection{Visible decay searches.}

At fixed-target experiments, the ALP can be produced through various processes, depending on its couplings to SM particles: from meson decays, through Primakoff scattering, through its mixing with the neutral pseudoscalar mesons, or via proton or electron Bremsstrahlung. Here we distinguish three different cases: 
\begin{itemize}
    \item For an ALP coupled to gluons through $\mathcal{O}_G$ in Eq.~\eqref{eq:GGdual}, constraints of this type are obtained from $K^\pm \to \pi^\pm \gamma\gamma$ measurements at NA62~\cite{NA62:2014ybm}, E949~\cite{E949:2005qiy}, $K_L \to \pi^0 \gamma\gamma$ at NA48/2~\cite{NA48:2002xke} and KTeV~\cite{KTeV:2008nqz}, and from $B\to K a,~a\to\gamma\gamma$ searches at BaBar~\cite{BaBar:2021ich}. We take these limits from Ref.~\cite{Goudzovski:2022vbt}. Additional limits come from searches for LLP decays into two photons at CHARM~\cite{CHARM:1985anb} (derived in Ref.~\cite{Jerhot:2022chi}). 
    \item For an ALP coupled through the $\mathcal{O}_{W}$ operator in Eq.~\ref{eq:Lag-EW}, significant constraints are obtained from E137 searches~\cite{Bjorken:1988as} (see e.g. Refs.~\cite{Dolan:2017osp,Dobrich:2019dxc,Essig:2010gu}), which we take from Ref.~\cite{Dolan:2017osp}, as well as from $K\to\pi\gamma\gamma$ (taken from Ref.~\cite{Goudzovski:2022vbt}) and from $B\to K a (a\to\gamma\gamma)$~\cite{BaBar:2021ich}. Additional bounds can be obtained from NA64, for an ALP produced in the forward direction through the Primakoff effect in the vicinity of a nucleus, recasting their search for ALPs decaying into two photons~\cite{NA64:2020qwq}. Finally, ALPs could be produced at LEP through an off-shell photon (for example, via $e^+ e^- \to \gamma^* \to \gamma a$) or in photon fusion ($e^+e^- \to e^+ e^- a$) and decay to two photons. The corresponding bounds are taken from Ref.~\cite{Jaeckel:2015jla}. 
    \item For an ALP coupled through the $\mathcal{O}_\phi$ operator in Eq.~\eqref{eq:Lag-EW}, relevant bounds are obtained recasting CHARM bounds on LLPs decaying into lepton pairs~\cite{CHARM:1985anb,CHARM:1983ayi}. Here we use the revised bounds obtained in Ref.~\cite{Dobrich:2018jyi} (see also Refs.~\cite{Essig:2010gu,Dolan:2014ska}). Significant constraints are also obtained from LHCb~\cite{LHCb:2015nkv,LHCb:2016awg}, for ALPs produced from $B$-meson decays (via $B \to K a$) and decaying within the detector into $\mu^+ \mu^-$. We take these from Refs.~\cite{Dobrich:2018jyi,Gavela:2019wzg}. Finally, in our previous work~\cite{Coloma:2022hlv} we obtained new bounds from a recast of a MicroBooNE search for $e^+e^-$ pairs from a long-lived particle pointing to the NuMI absorber~\cite{MicroBooNE:2021usw}. Here we also add to these the corresponding recast for a similar search into $\mu^+\mu^-$ pairs~\cite{MicroBooNE:2022ctm}, following the same methodology as in Ref.~\cite{Coloma:2022hlv}. We also note that the ArgoNeuT experiment has recently obtained a bound on axion-like particles for heavier masses (up to 700~MeV) decaying into $\mu^+ \mu^-$ pairs~\cite{ArgoNeuT:2022mrm}. These cannot however be easily recasted to the scenario considered here without a proper simulation of the meson fluxes in the NuMI target (in particular, from $\eta$ meson production), which lies beyond the scope of this work. 
    \item For an ALP coupled to right-handed quarks, Eq.~\eqref{eq:LagCharming}, relevant constraints can be obtained for CHARM~\cite{CHARM:1985anb}, using the null results from a search for $a\to \gamma\gamma$ as outlined in Ref.~\cite{Carmona:2021seb}. We rederive such constraint here, finding significant differences which we attribute to the different treatment of the decay width of the ALP as well as to a different simulation of the $D$ meson production. Specifically, we obtain the $D$-meson fluxes from Pythia (v8.3.07)~\cite{Bierlich:2022pfr} for a beam energy of 400~GeV, and compute the signal acceptance of the detector using the same methodology described in Sec.~\ref{Section:Production}, for a detector with a decay volume of $3\times3\times35 \mathrm{~m}^3$ located at $L_{det}=480$~m from the target. Our computation of the lifetime of the ALP, as well as the branching ratio for the $a\to\gamma\gamma$ decay, follows Ref.~\cite{Cheng:2021kjg}. 
\end{itemize}

Finally, we note that data from atmospheric neutrino oscillation experiments may provide relevant constraints as well. For example, in Refs.~\cite{Arguelles:2019ziu, Coloma:2019htx} a model-independent analysis of SK multi-ring data was able to set a constraint on HNL production  $BR(K\to N) \times BR(N \to e-\mathrm{like}) \lesssim 5\times 10^{-9}$ for values of $c\tau \sim \mathcal{O}(\mathrm{km})$. A similar analysis would also constrain the scenarios considered here, for ALP decays into $\gamma\gamma$ or $e^+e^-$.

\subsection{Invisible decay searches.} 

For very light masses, or for sufficiently small couplings, the ALPs may exit the detector without decaying (hence the term ``invisible decay''). Results for $K \to \nu\bar\nu$ or $B\to K \nu\bar\nu$ may then be reinterpreted as bounds on the branching ratios for $K\to \pi a$ and $B \to K a$~\cite{Essig:2010gu,Izaguirre:2016dfi,Gavela:2019wzg}. For heavier masses, while no dedicated search for $D \to \pi a$ exists, an indirect constraint may be obtained from the reinterpretation of $D^+ \to \tau^+ \nu, \tau^+ \to \pi^+ \bar\nu$ measurements, as pointed out in Refs.~\cite{MartinCamalich:2020dfe, Carmona:2021seb}.

The strongest limits on $K\to\pi a$ are obtained from the NA62 experiment~\cite{NA62:2020xlg,NA62:2020pwi}. However, the very competitive bounds from E787 \& E949~\cite{BNL-E949:2009dza} are comparable (and even dominate) in the region close to the pion mass. For $B \to K a$ decays, we use the constraint from Belle on $B^+ \to K^+ \nu\bar\nu$~\cite{Belle:2017oht}. Experimental limits on invisible ALP decays also arise from precision measurements of the pion momentum in $K \to \pi X$~\cite{Yamazaki:1984vg}. Here we draw from our previous work in Ref.~\cite{Coloma:2022hlv} where we derived the corresponding constraints for an ALP coupled through the electroweak operators in Eq.~\eqref{eq:Lag-EW}. For the gluon dominance scenario, we extract the relevant bounds from Ref.~\cite{Goudzovski:2022vbt}. A priori, similar bounds may also be derived for the Charming ALP scenario. However it can be seen that the leading contribution is obtained when the $c$-quark is running in the loop. In particular, following Refs.~\cite{Coloma:2022hlv, Bauer:2020jbp}, we find 
\begin{equation}
\label{eq:Kds-charming}
\bigg|\frac{[k_Q(\mu)]_{ds}}{V_{cd}^* V_{cs}}\bigg|_{\Lambda = 1~\mathrm{TeV}} \simeq 3.5 \times 10^{-6} \left[c_{u_R}(\Lambda)\right]_{22} \qquad (\mathrm{for}~\mathcal{O}_{u_R}). 
\end{equation}
Using this effective coupling, we re-derive here the corresponding constraints from $K\to \pi a$, following the procedure described in Ref.~\cite{Coloma:2022hlv}; however, while in Ref.~\cite{Coloma:2022hlv} the running was taken from $\Lambda = 1~\mathrm{TeV}$, here we take into account the relationship between $f_a$ and $\Lambda$, see Eq.~\eqref{eq:textureFN}. In other words, for a given value of $f_a$ the running of the coupling constants is carried out from $\Lambda = f_a / \epsilon$ down to $\mu = 2~\mathrm{GeV}$. 

For the charming ALP scenario, bounds on $D \to \pi a$ can be derived using CLEO~\cite{CLEO:2008ffk} or BESIII~\cite{BESIII:2019vhn} measurements of the $D \to \tau \nu$ decay. Their data is provided as a function of the missing mass squared in both cases, which can be used to derive a limit on BR($D\to\pi a$) as a function of $m_a^2$, as was done in Ref.~\cite{Carmona:2021seb} for CLEO. Here we decide to use BESIII data instead, as for CLEO we are not able to reproduce their SM fit with the information provided in Ref.~\cite{CLEO:2008ffk}. Specifically, we take the BESIII data from Fig.~3 in Ref.~\cite{BESIII:2019vhn}. We have checked that we are able to reproduce their result for the determination of $D\to\tau\nu$ in the SM within a reasonable degree of accuracy. Our best-fit curve leads to $\chi^2_\mathrm{min}/\mathrm{dof} \sim 29 / 28$, indicating a good compatibility with the data with a $p$-value of 0.4. We then use the fit to derive a constraint on the BR($D\to\pi a$) as a function of the mass of the ALP, see App.~\ref{app} for details.

Finally we note that additional bounds arise from searches for mono-photon signals at colliders~\cite{Izaguirre:2013uxa,Essig:2013vha,Izaguirre:2016dfi}. The best limits of this class are obtained using BaBar~\cite{BaBar:2008aby,BaBar:2017tiz} or LEP~\cite{L3:1997exg} data. Nevertheless, the resulting bounds are milder than the rest of the limits considered in this work and will not be shown here.

\subsection{Bounds from {\boldmath $D-\bar D$} mixing.}

Neutral meson particle-antiparticle pairs are allowed by the SM interactions to both oscillate and decay to lighter particles. Their oscillations are parametrized by a $2\times2$ unitary mixing matrix $M$. The mixing for $D^0-\bar D^0$ is very suppressed in the SM, leaving room to probe new physics that might induce such mixing. These bounds are relevant for the charming ALP scenario, since the non-vanishing $\left[c_{u_R}\right]_{21}$ coupling in Eq.~\eqref{eq:textureFN} leads to non-standard contributions in the form of effective operators at low energies involving four quarks. These contribute to the off-diagonal entry in the mixing matrix, $M_{12}$. Here we take the bounds derived in Ref.~\cite{Carmona:2021seb} using the mixing constraints on $M_{12}$ from Ref.~\cite{HFLAV:2019otj}. 

\subsection{Astrophysical bounds.} 

Within the mass range of interest here, the main constraints are derived from supernovae data. These can be further classified into three subcategories: (1) constraints obtained from the requirement that the energy loss induced by ALP emission does not exceed that of neutrino emission~\cite{Raffelt:1990yz,Raffelt:1987yt, Chang:2018rso,Caputo:2022rca,Lee:2018lcj}; (2) bounds coming from the visible photon signal resulting from the ALP burst~\cite{Jaeckel:2017tud,Diamond:2023scc}; and (3) limits obtained from the observation of low-luminosity core-collapse supernovae, which constrain the total energy deposition in the progenitor star from radiative ALP decays~\cite{Caputo:2022mah}. Additional limits may be derived from X-ray observations of neutron star mergers associated to gravitational waves~\cite{Diamond:2023cto}.

Here we take the limits on $c_W$ from Ref.~\cite{Goudzovski:2022vbt}, while for the charming ALP scenario we take the limits from Ref.~\cite{Carmona:2021seb}. Finally, while bounds derived on ALPs coupled to muons or electrons do exist in the literature\cite{Chang:2018rso,Lucente:2021hbp,Ferreira:2022xlw,Croon:2020lrf,Caputo:2021rux,Caputo:2022mah}, to the best of our knowledge none of these analyses may be easily recasted for our $\mathcal{O}_\phi$ operator (which induces couplings to electrons, muons and, via loop, to photons) within our mass range of interest. Consequently, no bounds from supernovae are included in this case. We stress, nonetheless, that astrophysical bounds are typically relevant for couplings much smaller than those within the expected sensitivity reach for DUNE, and are only shown here for completeness.

%%%%%%%%%%%%%%%%%%%%%%%%%%%%%%%%%%%%%%%%%%%%%%%%%%
\section{Targeting signals of LLP decays at the DUNE Near Detectors}
\label{sec:bg}
DUNE~\cite{DUNE:2020lwj} is an upcoming long-baseline neutrino oscillation experiment under construction in the United States. Once built, DUNE will consist of two state-of-the-art neutrino detectors exposed to the world’s most intense neutrino beam. The so-called \emph{near detector} (ND) will record neutrino interactions near the source of the beam, at the Fermi National Accelerator Laboratory (Fermilab), in Illinois. The much larger \emph{far detector} (FD) will be built at a depth of 1.5~km at the Sanford Underground Research Facility (SURF), in South Dakota, 1300~km away from Fermilab. DUNE has a very rich scientific program that includes the precision study of neutrino mixing \cite{DUNE:2020jqi, DUNE:2021mtg}, astroparticle physics \cite{DUNE:2020zfm} and searches for BSM phenomena \cite{DUNE:2020fgq}.

In this section, we discuss the prospects for detecting the decay of LLPs at the DUNE ND  \cite{DUNE:2021tad}, which will be built in a shallow underground hall located 574~m downstream from the neutrino beam origin. In the first phase of DUNE, the ND will consist of a liquid-argon Time Projection Chamber (TPC), ND-LAr, followed by a downstream muon spectrometer (the so-called \emph{temporary muon spectrometer}, or TMS). In the so-called Phase II of the experiment, TMS will be replaced with ND-GAr, a magnetized, high-pressure gaseous argon TPC surrounded by a calorimeter. We do not consider in our study the magnetized beam monitor known as SAND, which will be present in both phases of the experiment.

The search for LLP decays at the DUNE ND will suffer from a significant background from neutrino interactions, given the intensity of the LBNF neutrino beam. Considering this, ND-GAr appears to be the ideal detector for this kind of searches, given its large volume (proportional to the number of signal events) and low mass (proportional to the number of background events). However, ND-GAr will only be available as part of the upgrades contemplated for Phase II \cite{DUNE:2022aul}. Therefore, in this work we have also estimated the expected sensitivities for the ND-LAr detector, as it will start collecting data much sooner. Optimal selection cuts for ALP signals at ND-LAr and ND-GAr are generally different, due to their different features. Thus, our background analysis is performed for each detector separately.

%%%%%%%%%%%%%%%%%%%%%%%%%%%%%%%%%%%%%%%%%%%%%%%%%%
\subsection{Simulation}
We have used large simulation data sets and a basic event selection to estimate the signal efficiency and background rejection in the DUNE ND for the main LLP decay channels of our benchmark models. Each simulation event represents a single LLP decay or neutrino-argon interaction in the active volume of the detectors, thus ignoring possible pile-up effects (i.e. the cross-contamination of different interactions occurring in the same TPC event) since the ND design will be optimized to make them negligible \cite{DUNE:2021tad}. For simplicity, detector effects are not simulated, but we do take them into account in our study with the introduction of the typical detection thresholds and resolutions expected from ND-LAr and ND-GAr.

The first step in our simulation involves the generation of the LLP fluxes at the ND starting from the decays of their parent mesons. For kaon decays, we make use of publicly-available histogram files \cite{DUNE:2021cuw,ntuples} that contain the position and three-momentum distributions of the decay of mesons in the LBNF beamline as obtained with G4LBNF \cite{DUNE:2021cuw}, the official Geant4 simulation of the LBNF beamline from primary proton beam to hadron absorber. This simulation code includes kaon production in the target, along the 194~m-long decay pipe, and in the absorber at the end of the decay pipe, leading to a kaon production yield $Y_K = 0.75$ mesons/PoT. For the $D$ mesons, we use the same distributions as in Ref.~\cite{Coloma:2020lgy}. They were obtained using Pythia (version 8.2.44) \cite{Sjostrand:2014zea} to create a pool of events for proton collisions at various momenta, followed by a Geant4 simulation to predict proton inelastic interactions between the 120~GeV primary proton beam and the target. Doing this, the $D$-meson production yield is $Y_D = 1.2 \cdot 10^{-5}$~mesons/PoT. As for the luminosity considered, we take a total nominal exposure of $1.1\times10^{22}$~PoT, corresponding to 10~years of DUNE operation \cite{DUNE:2020jqi}. 

The parent mesons will decay to LLPs (as detailed in Sec.~\ref{Section:Production}), which will then propagate towards the DUNE ND, located 574~m away from the target. In our computation, we approximate the ND-LAr as a rectangular cuboid 7~m wide, 3~m high and 5~m deep, with an assumed fiducial volume that excludes 50~cm from the sides and upstream end and 150~cm from the downstream end \cite{DUNE:2021tad}. The active volume of the HPgTPC is a cylinder with a radius of 260~cm and a length of 500~cm; for the purpose of computing event rates, we define a fiducial volume by excluding the outer 37~cm of radius and 30~cm on each end of the cylinder \cite{DUNE:2021tad}.

We assume that the dominant background source in this search will be neutrino-argon interactions in the active volume of the TPCs. We estimate that other possible background sources (such as neutrino-rock interactions or cosmic muons) will be negligible in comparison, as the resulting events will not be aligned with the direction of the beam, in general. Using the GENIE neutrino Monte Carlo generator (version 3.2.0) \cite{Andreopoulos:2009rq} and the public DUNE flux histogram files \cite{DUNE:2021cuw,ntuples}, we have produced $2\times10^7$ $\nu_\mu$-Ar interactions. 

Below, we discuss the event selection we have devised for the four main ALP decay channels relevant for the models discussed in Sec.~\ref{Section:Production}. Table~\ref{tab:effbkg} summarizes our results. We have verified that our estimates do not change significantly for different LLP masses ($m_a$) in the relevant range for each channel. In this regard, it is worth noting that the reconstructed invariant mass of the LLP would provide a handle for the discrimination of signal and background that we have not exploited in our study.

%%%%%%%%%%%%%%%%%%%%%%%%%%%%%%%%%%%%%%%%%%%%%%%%%%
\begin{table}
\caption{\label{tab:effbkg} Signal efficiencies and background event rates for the different decay channels, before and after event selection according to the cuts discussed in the main text. Results are shown separately for the two DUNE near detectors considered. Background event rates are provided per year, and for the total fiducial volume considered for each detector. We highlight in bold type the large backgrounds expected for some of the decay channels, as well as the reduced LAr ND signal efficiencies for most decay channels considered.}
\begin{ruledtabular}
\begin{tabular}{clccrr}
%%%%%%%%%%%%%%%%%%%%%%%%%%%%%%%%%%%%%%%%
& Selection cut & \multicolumn{2}{c}{Signal efficiency} & \multicolumn{2}{c}{Background rate} \\ \cmidrule{3-4} \cmidrule{5-6}
& & ND-LAr & ND-GAr & ND-LAr & ND-GAr \\ \midrule
%%% mu+mu- %%%%%%%%%%%%%%%%%%%%%%%%%%%%%
\multirow{4}{*}{\rotatebox{90}{$\mu^+\mu^-$}} 
& Two $\mu$-like tracks only & 1.00 & 1.00 & 3545674 & 70656 \\
& PID $\mu$ and opposite charge sign & 0.40 & 1.00 & 6226 & 124 \\
& Transverse momentum $<0.125$~GeV/c & 0.40 & 0.99 & 99 & 2 \\
& Angle between muons $< 0.7$~rad & 0.40 & 0.94 & 0 & $ 0 $ \\ \midrule
%%% e+e- %%%%%%%%%%%%%%%%%%%%%%%%%%%%%%%
\multirow{2}{*}{\rotatebox{90}{$e^+e^-$}} 
& Two $e$-like tracks/showers & 0.10 & 1.00 & 9432 & 145 \\
& Reconstructed ALP direction & \textbf{0.10} & 0.99 & 180 & 15 \\ \midrule
%%% gg %%%%%%%%%%%%%%%%%%%%%%%%%%%%%%%%%
\multirow{3}{*}{\rotatebox{90}{$\gamma\gamma$}} 
& Two $\gamma$ showers only & 0.05 & 0.79 & 36276 & 14222 \\
& Reconstructed ALP direction & 0.05 & 0.79 & 6938 & \textbf{7923} \\
& Angle between $\gamma$ showers & \textbf{0.05} & --- & \textbf{1367} & --- \\ \midrule
%%% 3pi %%%%%%%%%%%%%%%%%%%%%%%%%%%%%%%%%
\multirow{4}{*}{\rotatebox{90}{$\pi^+\pi^-\pi^0$}} 
& Two $\mu$-like tracks, two $\gamma$ showers & 0.04 & 0.81 & 2030490 & 40462 \\
& PID $\pi^\pm$ and charge sign & 0.04 & 0.81 & 431035 & 8589 \\
& Transverse momentum $<0.2$~GeV/c & 0.04 & 0.79 & 17182 & 342 \\
& Angle between pions $< 0.15$~rad & \textbf{0.04} & 0.69 & \textbf{946} & 19 \\ 
\end{tabular}
\end{ruledtabular}
\end{table}
%%%%%%%%%%%%%%%%%%%%%%%%%%%%%%%%%%%%%%%%%%%%%%%%%%

%%%%%%%%%%%%%%%%%%%%%%%%%%%%%%%%%%%%%%%%%%%%%%%%%%
\subsection{Event selection: $\mu^+\mu^-$ decay channel}
A priori, the most important background source for the di-muon decay channel is $\nu_\mu$ charged-current events with charged pions, as it is relatively easy to confuse muon and pion tracks due to their similar stopping power ($d\mathrm{E}/d\mathrm{x}$) in argon. About 38\% of the $\nu_\mu$-Ar interactions have a charged muon and a charged pion above threshold in the final state, with an expected rate of the order of $6\times10^5$ events per ton-year at the DUNE ND. Actual di-muon events from charged-current charm production only represent less than one percent of the total background events. 

We start our event selection requiring candidate signal events to have only two $\mu$-like tracks (i.e., $\mu^\pm$ or $\pi^\pm$) above threshold. This allows the rejection of background events with hadronic activity near the interaction vertex. We consider a proton detection threshold of 40~MeV for the ND-LAr and of 5~MeV for the ND-GAr \cite{DUNE:2021tad}, obtaining similar rejection factors for both detectors: about 0.3\% of the initial events meet the above criterion. From this point on, each detector requires slightly different considerations, discussed next.

\begin{figure}
\centering
\includegraphics[width=0.495\linewidth, trim={0 20 0 35},clip]{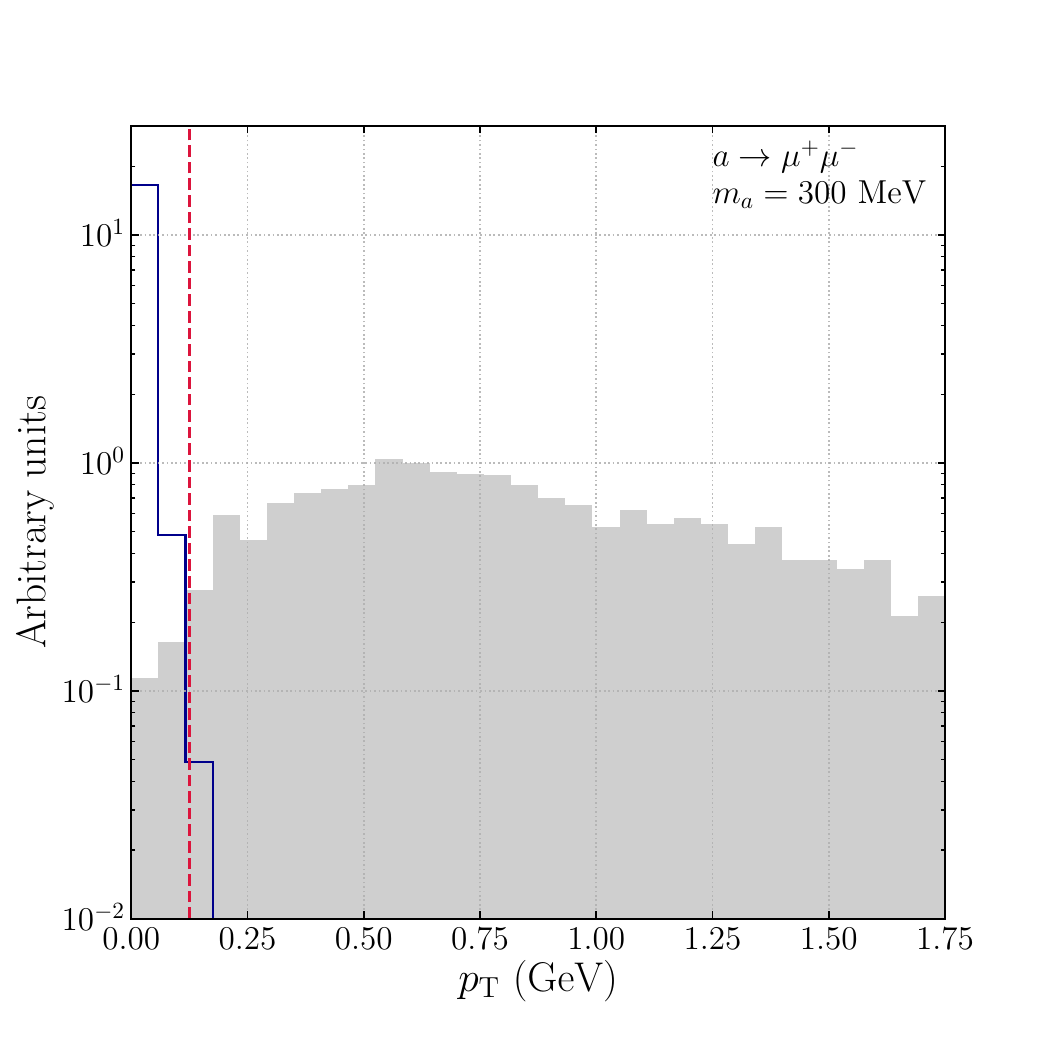}
\includegraphics[width=0.495\linewidth, trim={0 20 0 35},clip]{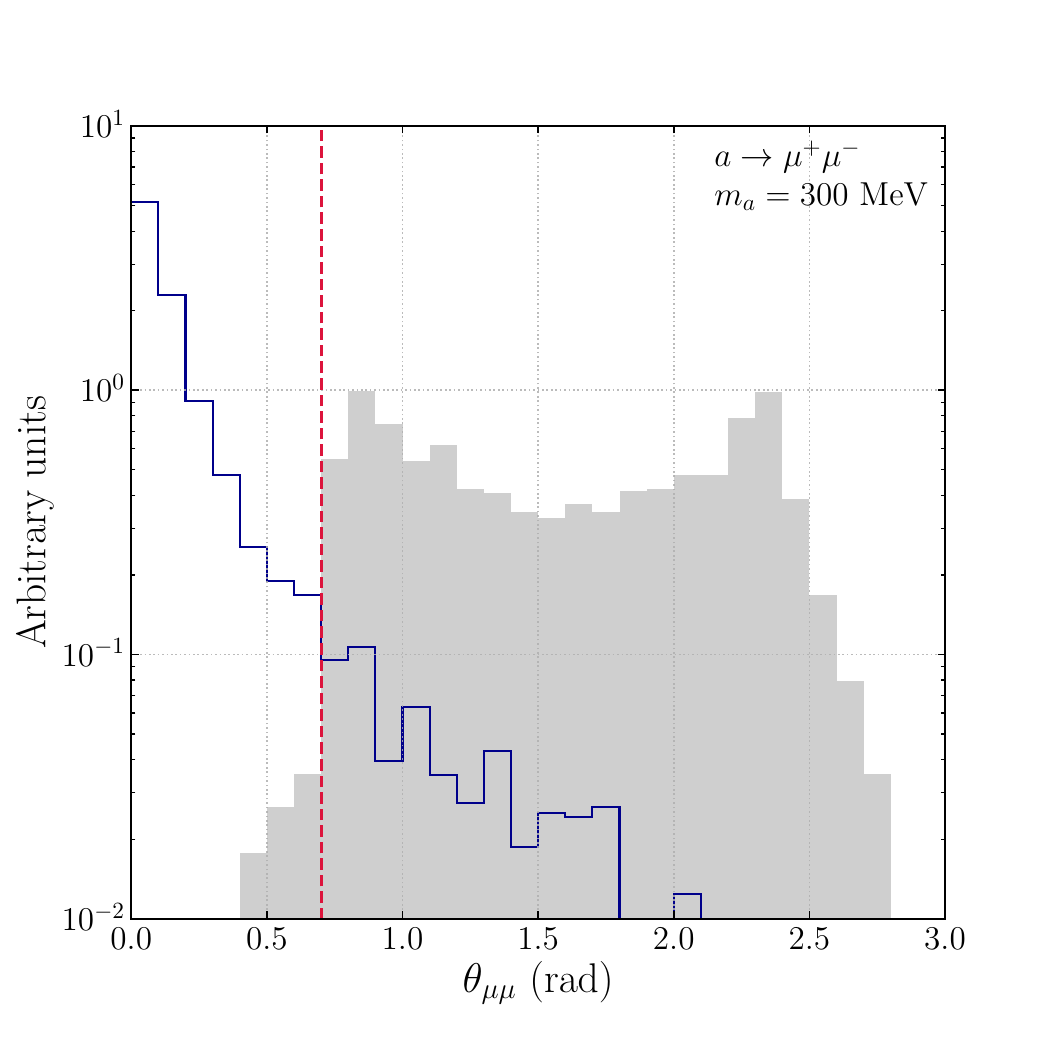}
\includegraphics[width=0.495\linewidth, trim={0 20 0 35},clip]{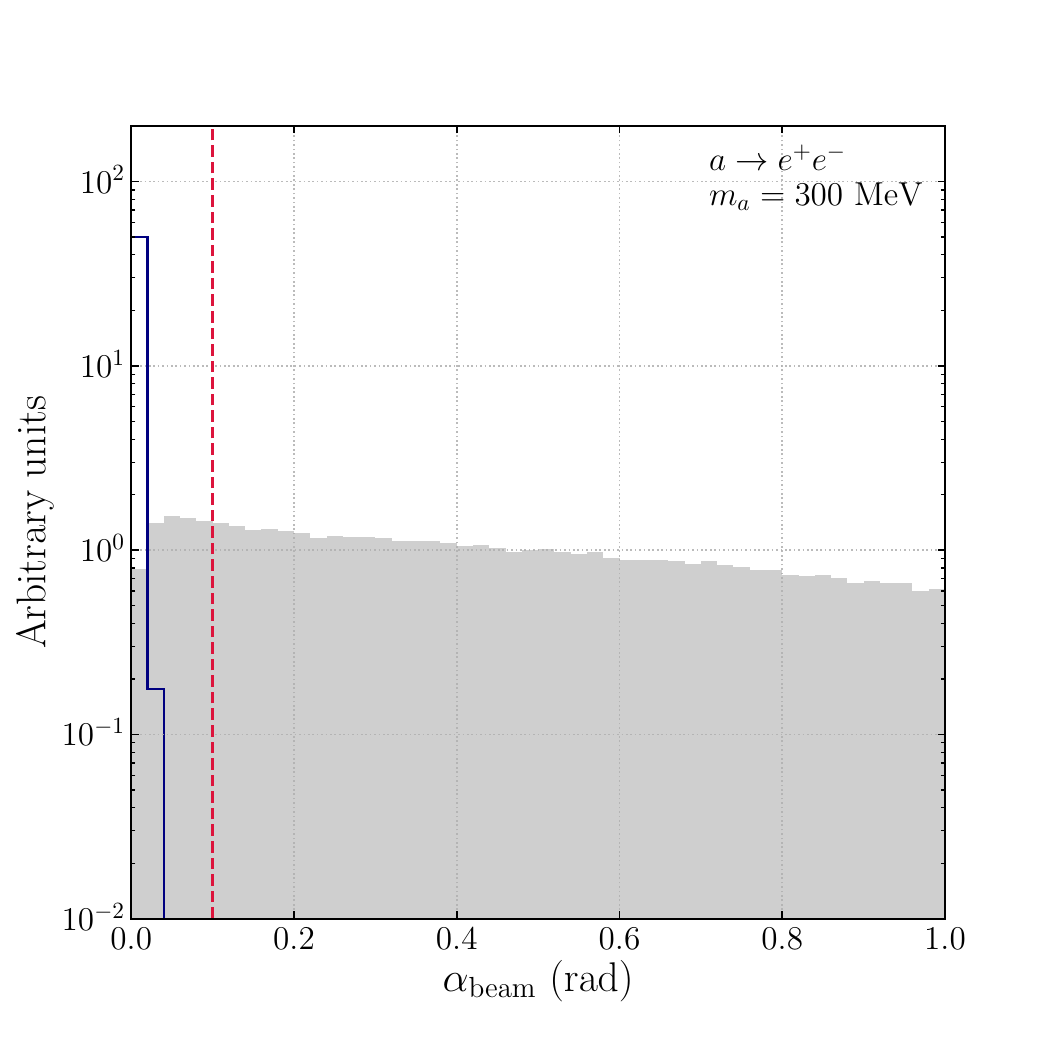}
\includegraphics[width=0.495\linewidth, trim={0 20 0 35},clip]{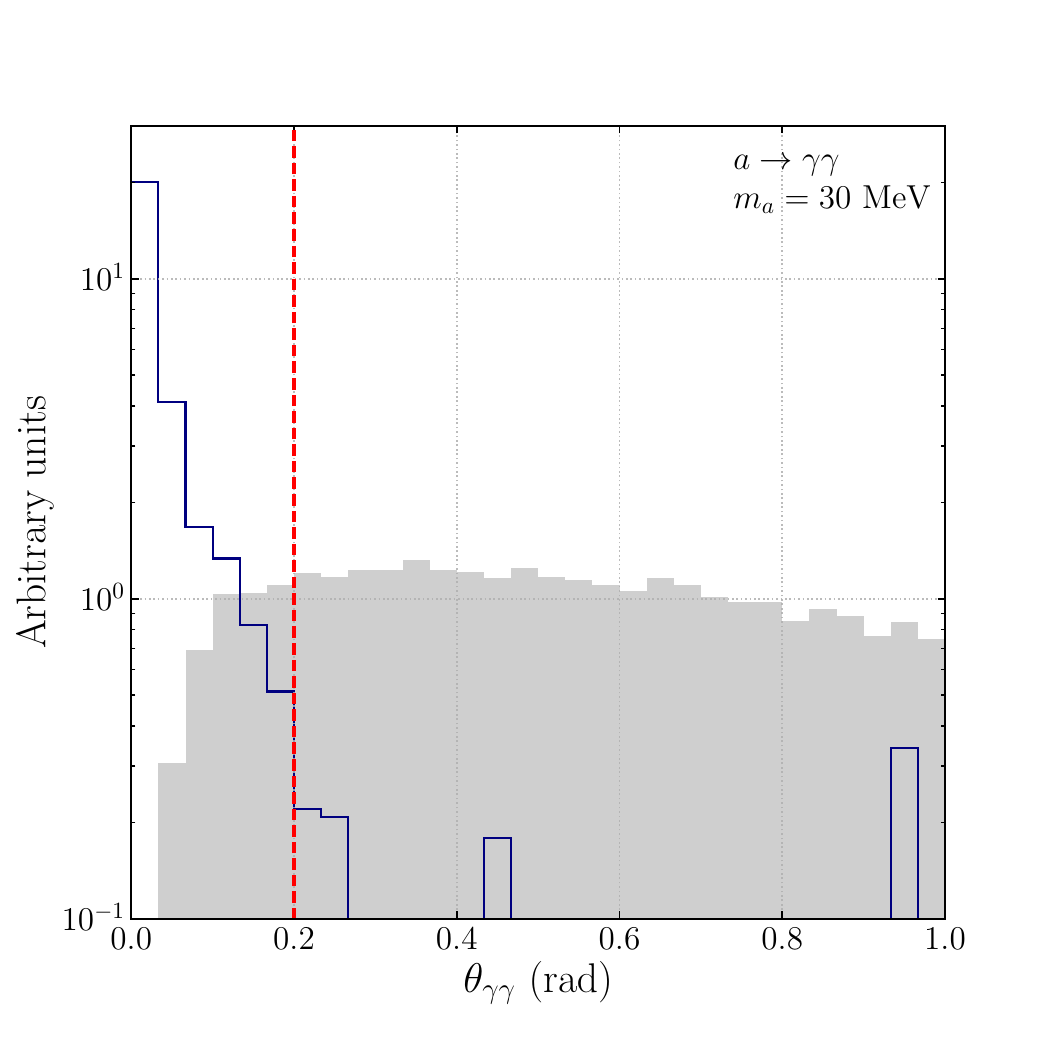}
\includegraphics[width=0.495\linewidth, trim={0 20 0 35},clip]{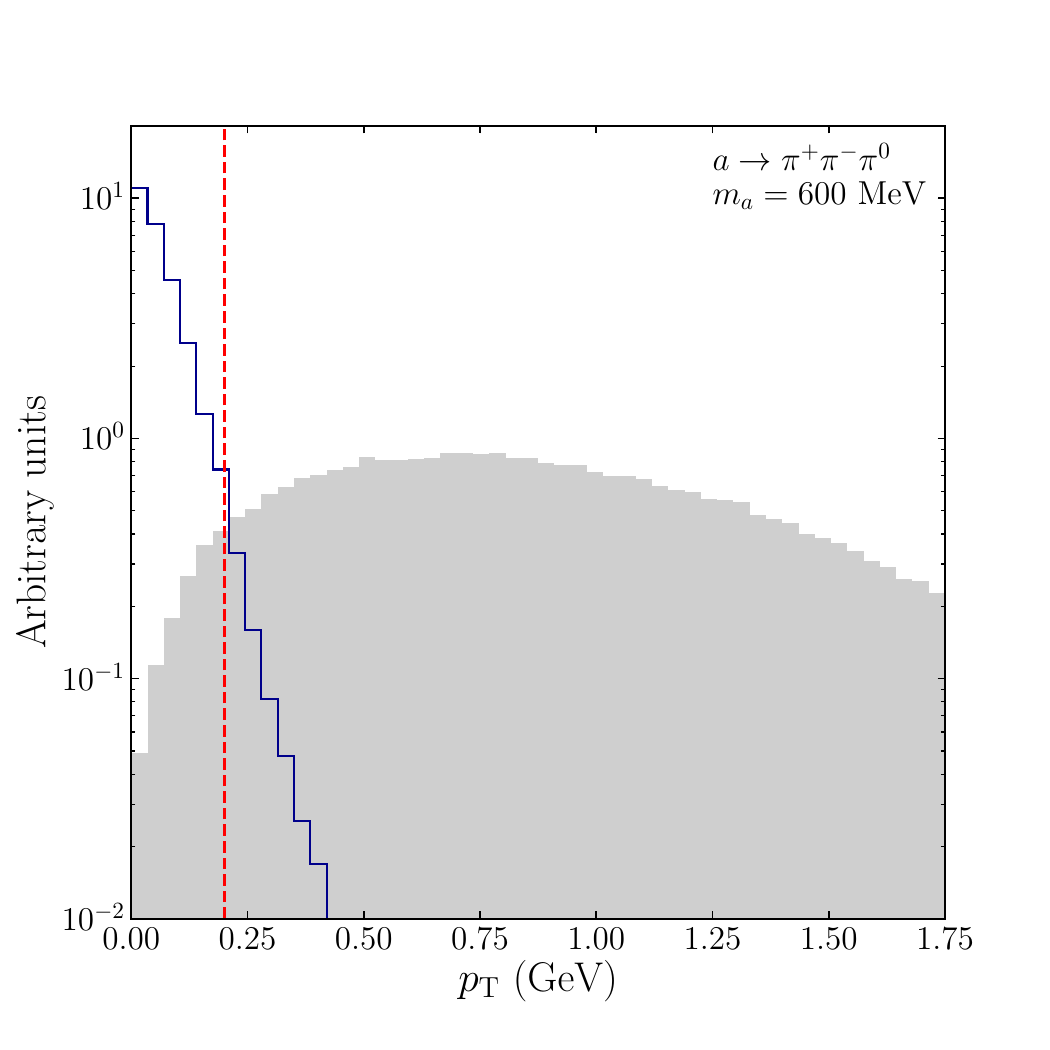}
\includegraphics[width=0.495\linewidth, trim={0 20 0 35},clip]{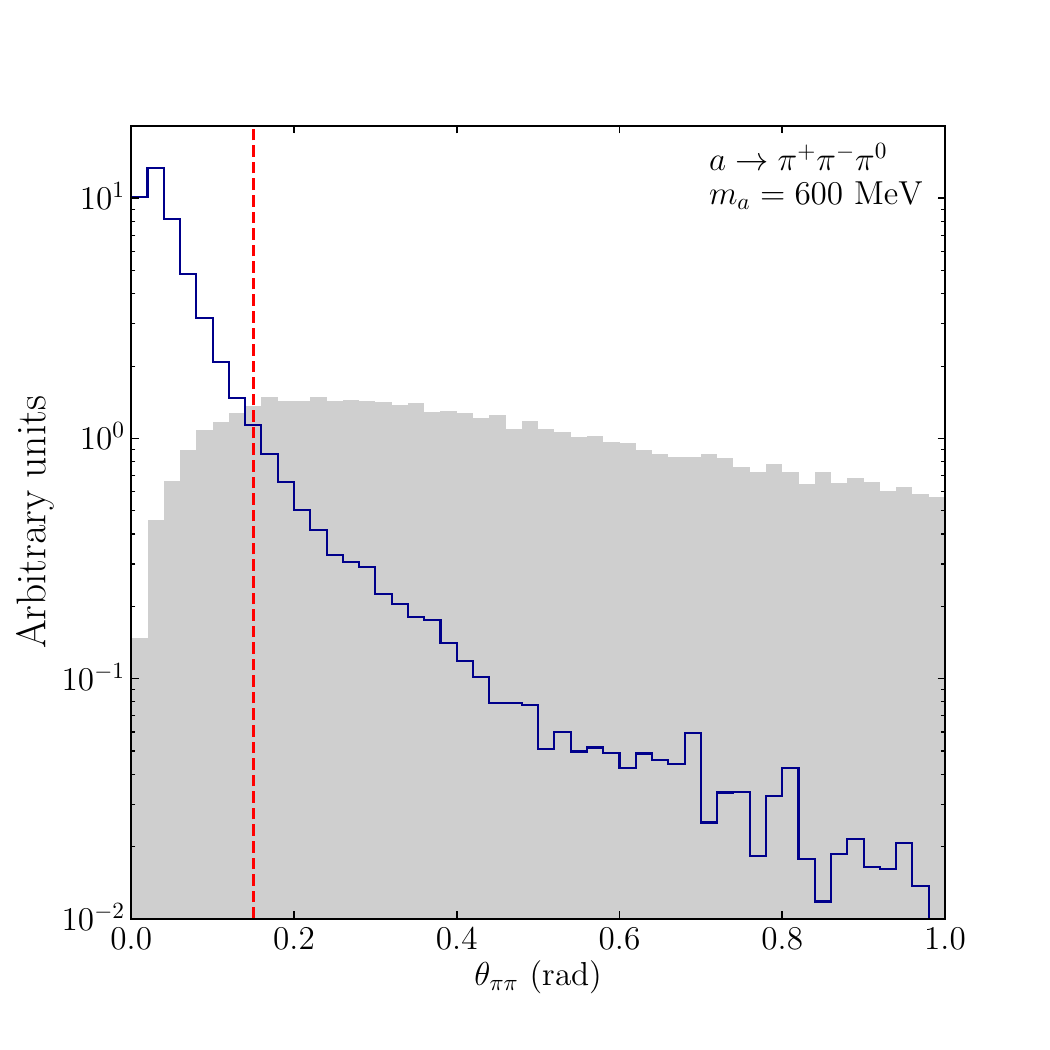}
\caption{Distributions for signal (blue line) and background (grey solid histogram) of the kinematic quantities used in our event selection. The vertical dashed red line indicates the selection cut value used. All histograms are area normalized. Top row: transverse momentum (left panel) and angle between tracks (right) for $\mu^+\mu^-$ candidate ALP-decay events in the DUNE ND-GAr. Central row, left: angle with respect to the beam direction of $e^+e^-$ events. Central row, right: angle between the two photon showers for $\gamma\gamma$ candidate events in ND-LAr. Bottom row: transverse momentum (left panel) and angle between the charged-pion tracks of $\pi^+\pi^-\pi^0$ candidate events.}
\label{fig:selection_cuts}
\end{figure}

In ND-GAr, the TPC combined with the electromagnetic calorimeter (ECAL) and the muon identification system that surround it will provide superb $\mu/\pi$ separation capabilities, reaching 100\% purity in the identification of muons for a wide range of momenta \cite{DUNE:2021tad}. Moreover, the magnetic field in ND-GAr will allow the measurement of the charge sign of muons. Thanks to these capabilities, we can reduce the initial sample of background events by a factor $6\times10^{-6}$ for, essentially, perfect signal efficiency. Lastly, we can further reduce the background sample taking into account the particular kinematics of signal events (see Fig.~\ref{fig:selection_cuts}, top row): 
\begin{enumerate*}[label=(\roman*)]
    \item The reconstructed LLP transverse momentum should be low; that is, the LLP trajectory points back in the direction of the target.
    \item The muons in signal events are highly boosted, and thus the angle between them should be small.
\end{enumerate*}
We assume that a momentum resolution of the order of 5\% or better and angular resolution of the order of a few degrees can be achieved in ND-GAr for momenta up to $10~\mathrm{GeV/c}$ \cite{DUNE:2021tad}. Overall, these cuts let us achieve a background rejection in excess of $10^7$ (see Tab.~\ref{tab:effbkg}), resulting in a background-free search in 10~years of data taking.

In the case of ND-LAr, the detector will not be able to fully contain high-energy muons or measure lepton charge, but the downstream spectrometers (TMS in the first phase of DUNE, and ND-GAr in the second one) will measure the charge sign and three-momentum of the muons that enter them. Events with muon kinetic energies below 1~GeV will be contained within ND-LAr, while events with higher energy muons traveling within 20~degrees of the beam direction will exit ND-LAr and enter the spectrometer \cite{DUNE:2021tad}. TMS will only be able to measure muons up to $\sim6~\mathrm{GeV/c}$ before they range out, corresponding to 40\% of our LLP decays.\footnote{ND-GAr, which will use the curvature in the magnetic field to reconstruct the momentum, will be able to reconstruct muon tracks well up to $10~\mathrm{GeV/c}$ and beyond, improving the selection efficiency to 54\% of the decays.} We will assume as well that the combination of the $d\mathrm{E}/d\mathrm{x}$ measurement in ND-LAr plus the $\mu/\pi$ separation capabilities of the TMS ---\thinspace pions will interact inelastically in the steel layers of TMS with high probability, while muons will behave as minimum ionizing particle\thinspace--- will be enough to reach essentially perfect purity in the identification of muons, such as in the ND-GAr. Finally, the two kinematical cuts described above are applied, achieving a background-free search in 10~years of data taking. As a point of comparison, the analysis described in Ref.~\cite{Altmannshofer:2019zhy} for the identification of di-muon neutrino trident events achieved a background suppresion of 6 orders of magnitude using exclusively kinematical cuts in ND-LAr.

%%%%%%%%%%%%%%%%%%%%%%%%%%%%%%%%%%%%%%%%%%%%%%%%%%
%%%%%%%%%%%%%%%%%%%%%%%%%%%%%%%%%%%%%%%%%%%%%%%%%%

%%%%%%%%%%%%%%%%%%%%%%%%%%%%%%%%%%%%%%%%%%%%%%%%%%
\subsection{Event selection: $e^+e^-$ decay channel}
The dominant background source for this decay channel is $\nu_\mu$ neutral-current single-pion (NC$\pi^0$) events. They can be mistaken as an $e^+e^-$ signal if only one of the photons from the $\pi^0\to\gamma\gamma$ decay ---\thinspace with a branching ratio close to 99\% \cite{Workman:2022ynf}\thinspace--- converts in the detector, and no visible hadronic activity occurs at the vertex of the neutrino interaction. Less than 1\% of the $\nu_\mu$-Ar interactions result in a single $\pi^0$, corresponding to an expected rate of about $1.5\times10^4$ events per ton-year at the DUNE ND.

In LAr, the attenuation length for gamma rays in the energy range from 100~MeV to 10~GeV is of the order of 20~cm \cite{XCOM}. This limits considerably the probability that only one of the two photons from a $\pi^0$ decay interacts in ND-LAr. Through Monte Carlo simulation, we have estimated that probability to be 1.3\%. Conversely, in high-pressure argon gas, the attenuation length for gammas is well above 10~m. Therefore, in most cases, both gammas from the $\pi^0$ decay will escape the TPC, interacting in the ECAL; only in about 1\% of the events the decay will result in a single-gamma conversion in the GAr, with no associated activity in the ECAL \cite{DUNE:2021tad,Berryman:2019dme}.

Lastly, pair-conversion events in both ND-LAr and ND-GAr can be suppressed requiring the reconstructed direction of the LLP to be aligned with the neutrino beam (see Fig.~\ref{fig:selection_cuts}, middle left panel). Background events, in contrast, will follow a nearly isotropical distribution, as the neutral pions will be coming from neutrino-nucleus interactions. A rather conservative cut of 100~mrad (about $5^\circ$) for both ND-GAr and ND-LAr suppresses the background events by an order of magnitude.

For ND-LAr, as far as we are aware, there is no estimation yet of the reconstruction efficiency of high-energy electron showers. Therefore, taking into account the significant depth of the showers (more than 2~m for an electron of 1~GeV) and the busy DUNE ND environment, we conservatively assume an efficiency of 10\%.

%%%%%%%%%%%%%%%%%%%%%%%%%%%%%%%%%%%%%%%%%%%%%%%%%%
\subsection{Event selection: $\gamma\gamma$ decay channel}
The most important background for this channel is, like in the previous case, the decay of neutral pions resulting from $\nu_\mu$ neutral-current interactions. 

In the case of ND-GAr, we assume that these events can only be reconstructed with good efficiency and high purity if both gamma rays are detected in the ECAL \cite{DUNE:2021tad}. Nonetheless, this calorimeter will not be able to contain completely the electromagnetic showers from high-energy photons up to 10~GeV, as it will only be 8--12 radiation-lengths deep \cite{DUNE:2021tad}. The impact of this shower leakage on the energy and angular resolutions can only be fully understood with a detailed simulation, beyond the scope of this paper. We will only assume here that background events can be suppressed taking into account that signal events should point back in the direction of the neutrino beam. As the pointing accuracy of photon showers reconstructed in the ND-GAr ECAL is low (the resolution for photons of 1~GeV is about $10^\circ$), a selection cut on the opening angle between the two photons has limited impact and, hence, we do not consider it.

In ND-LAr, full reconstruction of both gammas requires the containment of a significant fraction of the two electromagnetic showers within the fiducial volume of the detector. Following Ref.~\cite{MicroBooNE:2022zhr}, we estimate the reconstruction efficiency to be 5\% in our energy range of interest. As ND-LAr should be able to reconstruct with accuracy of a few degrees the direction of the showers, the kinematical cut on the LLP direction results more effective than in the previous case, and we can also discriminate between signal and background applying a cut on the angle between the two photons (see Fig.~\ref{fig:selection_cuts}, middle right panel).

In this channel, the invariant mass of the di-photon system could be used effectively for background rejection, as the distribution for background events would peak around the $\pi^0$ mass (conversely, this limits as well the sensitivity to LLPs with mass in that energy region). As this selection cut requires a detailed understanding of the invariant mass resolution of the detectors, we are not applying it in this analysis.

%%%%%%%%%%%%%%%%%%%%%%%%%%%%%%%%%%%%%%%%%%%%%%%%%%
\subsection{Event selection: $\pi^+\pi^-\pi^0$ decay channel}
The most important background source for this decay channel is $\nu_\mu$ neutral-current events with multi-pion production, with an expected rate of the order of 39000 events per ton-year.
The same arguments given in previous sections for the reconstruction of $\mu/\pi$ tracks and the two photon showers from the $\pi^0$ decay apply here. Background events can be suppressed with selection cuts on the transverse momentum and angle between the two charged-pion tracks (see Fig.~\ref{fig:selection_cuts}, bottom row).
 
%%%%%%%%%%%%%%%%%%%%%%%%%%%%%%%%%%%%%%%%%%%%%%%%%%
\section{Results}
\label{sec:results}
To compute the expected sensitivity to the Wilson coefficients in Eq.~\eqref{eq:Lag-EW}, or to the value of $f_a$ in Eqs.~\eqref{eq:GGdual} and~\eqref{eq:LagCharming}, we perform, for each detection channel, an unbinned Gaussian $\chi^2$ analysis that takes into account the expected background event rates as outlined in Sec.~\ref{sec:bg} (see Table~\ref{tab:effbkg}). As the backgrounds stem mainly from neutrino neutral-current interactions in the detector, we include one nuisance parameter $\left(\xi \right)$ in order to account for systematic uncertainties affecting their overall normalization. This is done with the pull-method approach~\cite{Fogli_2002} taking a prior uncertainty on the background, $\sigma_{bg}=20\%$, to account for the large uncertainties coming from the corresponding cross section (see, e.g., Refs.~\cite{ArgoNeuT:2015ldo, MicroBooNE:2022zhr}). Thus, for a given decay channel $ch$ with an expected non-zero background rate $N_{bg, ch}$, we define our $\chi^2$ simply as
\begin{equation}
\label{eq:chi2}
\chi^2=\min _{\xi_{bg}}\left\{\left(\frac{T_{ch} (\lbrace\Theta,\xi\rbrace) - O_{ch} }{\sigma}\right)^2+\frac{\xi^2}{\sigma_{bg}^2}\right\} \, ,
\end{equation}
where $T_{ch}$ is the total expected event rate including both signal and background events, while $O_{ch}$ corresponds to the assumed observed events, and $\lbrace \Theta \rbrace$ stands for the parameters of the model. We take the observed events as expectation in the absence of a BSM signal, $O_{ch} = N_{bg, ch}$, and the associated statistical error as $\sigma = \sqrt{O_{ch}}$. The predicted event rates read:
\begin{equation}
T_{ch}(\lbrace \Theta, \xi \rbrace) = N_{dec, ch}(\lbrace \Theta \rbrace) 
+ (1+\xi) N_{bg, ch} \, .
\end{equation}
Note that the $\chi^2$ definition above, in terms of total event rates, will typically lead to conservative results: an improvement in sensitivity may be obtained for a binned analysis that takes into account the different distributions of the signal versus the background in the kinematic variables of interest, which we leave for future work.

Our sensitivity regions are obtained taking the corresponding $\chi^2$ cut at a given confidence level (C.L.), for 1 degree of freedom (d.o.f.). Thus, they can be interpreted as the upper limit that DUNE would be able to set on a given parameter, for an ALP with mass $m_a$, in the absence of a BSM physics signal. Finally, for the channels without SM background, we follow the Feldman-Cousins prescription~\cite{Feldman:1997qc} and require $N_{dec, ch} > 2.44$ for limits at 90\% C.L.

%%%%%%%%%%%%%%%%%%%%%%%%%%%%%%%%%%%%%%%%%%%%%%%%%%
\subsection{Model-independent sensitivity limits}
\label{sec:modelindep}
Using the $\chi^2$ analysis just outlined, we first perform a model-independent sensitivity analysis. If we assume that the production branching ratio and the lifetime of the ALPs are independent, the number of decays approximately depends as in Eq.~(\ref{eq:Ndec-approx}). This allows to derive a sensitivity limit on the product of the production and decay branching ratios as a function of $c\tau_a / m_a$, shown in Fig.~\ref{fig:bands}. Mild differences are obtained for different masses, however, induced by the dependence of the detector acceptance on $m_a$ (which affects the boost of the particles to the lab frame). Here we follow the same approach as in Ref.~\cite{Coloma:2023adi} and provide our limits as bands, where the width indicates the variation in the obtained limit when the mass of the LLP is varied between 10~MeV (upper edge of each band) and up to the production threshold in each case (lower edge of the band). In Fig.~\ref{fig:bands} we show two sets of bands, depending on the parent meson: kaons (blue) and $D$ mesons (red). Moreover, for each parent meson, we show results for $a\to \gamma\gamma$ using the ND-GAr, computed taking the corresponding background event rates from Table~\ref{tab:effbkg}, as well as the limiting sensitivity in the background-free case (which would only be applicable for decays into $\mu^+ \mu^-$). Note, however, that the {upper edge of each band corresponds to $m_a = 10~\mathrm{MeV}$, for which the decays into $\mu^+ \mu^-$ or multi-pion final states are not kinematically accessible. The best sensitivity for $a\to\mu^+\mu^-$ searches is indicated by the dashed lines in each case, corresponding to $m_a \sim 2 m_\mu$. 

%%%%%%%%%%%%%%%%%%%%%%%%%%%%%%%%%%%
\begin{figure}
\centering
\includegraphics[width=0.75\textwidth]{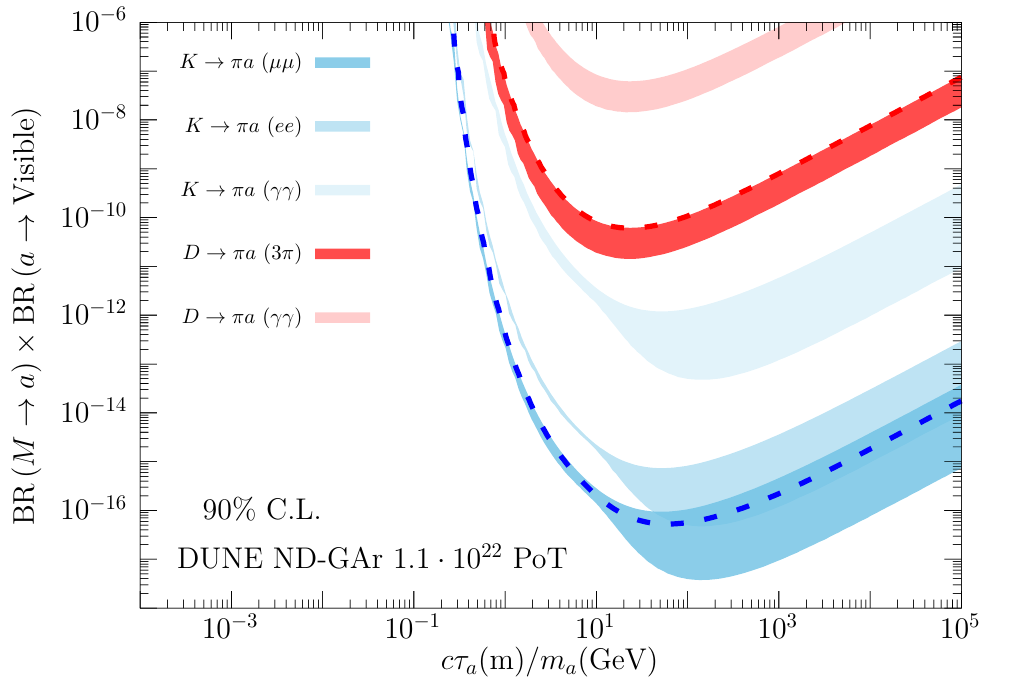}
\caption{\label{fig:bands} 
Expected sensitivity to LLPs in the
model-independent scenario, assuming their branching ratios
and lifetime are completely uncorrelated.
In the absence of a new physics signal, DUNE is expected to disfavor the region above each line at 90\% confidence
level (C.L.). The width of the bands indicates the variation
in our results when the mass of the LLP is varied between
10~MeV (upper edge of each band) up to the production threshold in each case; see text for details. The dashed lines correspond to $m_a \sim 2 m_\mu$, and therefore indicate the best sensitivity for searches of $a \to \mu^+\mu^-$. Blue (red) bands show the results for LLP produced from kaon ($D$-meson) two-body decays. For each parent meson, the different bands are obtained assuming different background rates; see main text for details.}
\end{figure}
%%%%%%%%%%%%%%%%%%%%%%%%%%%%%%%%%%%

In order to compare to current limits in the literature it is convenient to pick a specific mass. This exercise is done in Fig.~\ref{fig:br_vs_ctau}, where the show the DUNE sensitivity limits compared to previous constraints, for three representative masses: $m_a = 50$~MeV (left) and 300~MeV (center), where the strongest limits at DUNE would be obtained for LLP produced in $K$ decays; and $m_a=1.2$~GeV (right), which could be probed at DUNE only if the LLP is produced from $D$ decays. 

%%%%%%%%%%%%%%%%%%%%%%%%%%%%%%%%%%%
\begin{figure}
\centering
  \includegraphics[width=\textwidth, trim=0 30 0 0]{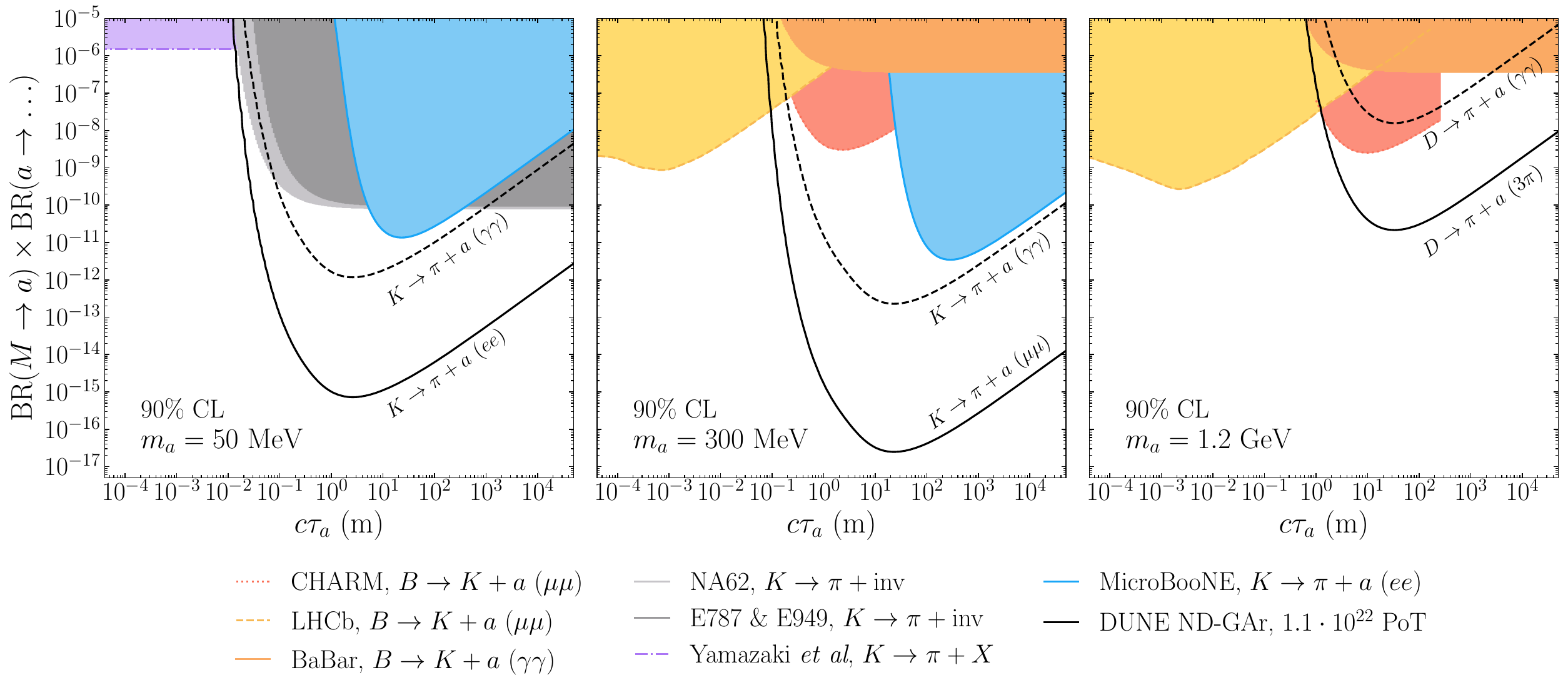}
\caption{\label{fig:br_vs_ctau} 
Expected sensitivity to LLPs as a function of their lifetime. The different panels show the results for $m_a=50$~MeV (left), 300~MeV (center) and 1.2~GeV (right), as indicated. Colored regions are disfavored by present constraints from searches for invisible and visible LLP decays, see Sec.~\ref{sec:bounds}. The different lines for DUNE (solid, dashed, dotted) are obtained for different background assumptions, depending on the final state being considered ($a\to\gamma\gamma, a\to \ell^+\ell^-, a\to \pi^0\pi^+\pi^-$), see Sec.~\ref{sec:bg}. In the case of the right panel, the limit for $a\to e e$ would coincide with the one shown for $a\to3\pi$, as backgrounds would be similar for both final states, while the limit for $a\to\mu\mu$ would be slightly better since it would be background free. Present constraints are indicated by the shaded areas and correspond to 90\% C.L., with the exception of CHARM and LHCb (shown at 95\% C.L.).}
\end{figure}
%%%%%%%%%%%%%%%%%%%%%%%%%%%%%%%%%%%

As shown in the figure, the sensitivity limits at DUNE depend heavily on the final state the LLP decays into, due to the different backgrounds expected. In particular, for ALP decays into $\gamma\gamma$ the analysis would be affected by large backgrounds, leading to a limit that is considerably worse than in the background-free scenario. This should be taken into consideration when comparing our results with similar studies in the literature obtained neglecting the effect of backgrounds. However, we see that DUNE is expected to improve over present constraints for masses below the kaon mass even in the less favorable case where the LLP decays into photons. In the case of decays into lepton pairs, DUNE is expected to reach values of $\mathrm{BR}(M \to a) \times \mathrm{BR}(a\to \ell^+\ell^-) \sim 10^{-17}$ for optimal values of the LLP lifetime\footnote{Model-independent DUNE sensitivities have been recently computed in Ref.~\cite{Batell:2023mdn}, where the authors considered LLPs decaying into $e^+e^-$. Our results, when rescaled according to the same number of PoT and assuming negligible backgrounds, show a reasonable agreement.}, many orders of magnitude below current constraints. In case of a heavier LLP (right panel), the expected limits suffer from a significant reduction in the production of the parent meson, and we find that the limit for an ALP that decays purely into photons is not as strong as the one obtained from CHARM data (salmon shaded regions, with dotted edges). However, we see that DUNE will be able to considerably improve over current limits if the ALP decays predominantly into lepton pairs or into $3\pi$, which would be background-free at the ND-GAr.  

\subsection{Sensitivity limits for specific scenarios}

The computation of model-independent sensitivity limits (Figs.~\ref{fig:bands} and~\ref{fig:br_vs_ctau}) is useful since allows our results to be easily recasted to other scenarios. However, one should bear in mind that in particular models, correlations may arise between the production and decay of the LLP if they depend on the same set of model parameters. This changes the relative importance of different sets of constraints, as these may be optimal for different values of the lifetime of the LLP. Moreover, while in Sec.~\ref{sec:modelindep} the constraints shown are obtained for searches for invisible or visible LLP decays, in specific scenarios additional bounds arise (e.g. from SN1987A or meson mixing, see Sec.~\ref{sec:bounds}). Thus, in the rest of this section we evaluate the sensitivity of DUNE for the benchmark scenarios considered in Sec.~\ref{Section:Production}, as illustrative examples. 

\subsubsection{Gluon dominance}

As outlined in Sec.~\ref{Section:Production}, the main production mechanism in this scenario is through ALP mixing with neutral pseudoscalar mesons (for ALP masses below $\sim 1~\mathrm{GeV}$) and gluon fusion (for higher masses). Once produced, if the ALPs are sufficiently long-lived they will reach the DUNE detectors before decaying into either a pair of photons or multi-pion final states, see Fig.~\ref{fig:BR-cgg}. Here we revisit the results previously computed in Refs.~\cite{Kelly:2020dda,Jerhot:2022chi} for this scenario, in light of our background estimates in Sec.~\ref{sec:bg} and the refined computation of the relevant decay widths in Ref.~\cite{Cheng:2021kjg}. Our results are shown in Fig.~\ref{fig:dune-sens-cgg}, where we show separately the expected sensitivity regions for a search for $a\to\gamma\gamma$ (red lines) and $a\to 3\pi$ (blue lines). Moreover, due to the different background rejection capabilities, we show separately the results for ND-LAr (dashed lines) and ND-GAr (solid lines). As can be seen from the figure, DUNE is expected to improve over current limits (shown by the shaded gray regions) only for ALP masses large enough to have a significant branching ratio to multi-pion final states, whereas for lighter ALP masses the sensitivity is affected by the large backgrounds expected for the di-photon channel. Also, note that even though the expected background for $a\to\gamma\gamma$ is higher for ND-GAr than for ND-LAr, its much higher signal efficiency leads to a better performance thanks to the higher signal-to-background ratio. The vertical dotted lines indicate the masses of the neutral pseudoscalar mesons, where the large mixing with the ALP leads to sharp features in our regions (see also Fig.~\ref{fig:BR-cgg}).

%%%%%%%%%%%%%%%%%%%%%
\begin{figure}[ht!]
\begin{center}
  \includegraphics[width=0.8\textwidth]{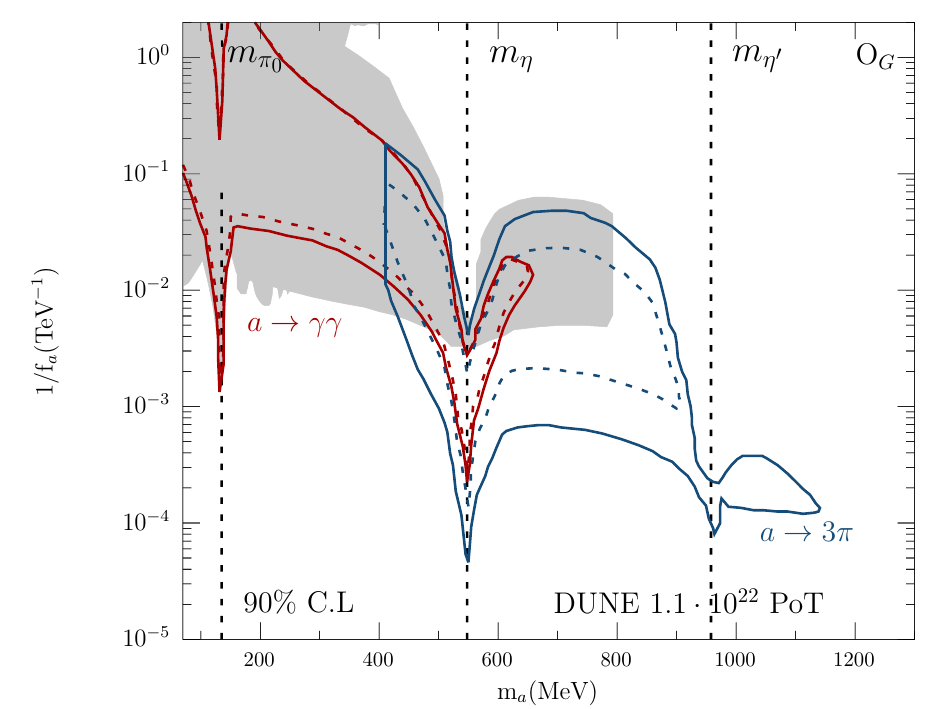}
\end{center}
\caption{\label{fig:dune-sens-cgg} DUNE sensitivity projections for the gluon dominance scenario, Eq.~\eqref{eq:GGdual}. Shaded regions are disfavored by present experiments. Red (blue) lines show the expected DUNE sensitivity (at 90\% CL) for a search into $a\to\gamma\gamma$ and $a\to 3\pi$ final states. Results are shown separately for the ND-LAr (dashed lines) and ND-GAr (solid lines). 
}
\end{figure}
%%%%%%%%%%%%%%%%%%%%%

\subsubsection{ALPs coupled through EW operators}

As outlined in Sec.~\ref{Section:Production}, the main production mechanism at DUNE for this scenario is through kaon decays, $K\to \pi a$. Once produced, the ALP may decay into three different final states: $e^{+} e^{-}$, $\mu^{+} \mu^{-}$ and $\gamma\gamma$, while hadronic modes are not allowed in this mass range (ALP decays $a\to \pi \pi$ and $a\to\pi^0\gamma$ are forbidden by $CP$ and $C$, respectively, and the decay into three pions is kinematically not allowed in this mass window). 
%%%%%%%%%%%%%%%%%%%%%
\begin{figure}[ht!]
\begin{center}
  \includegraphics[width=0.99\textwidth]{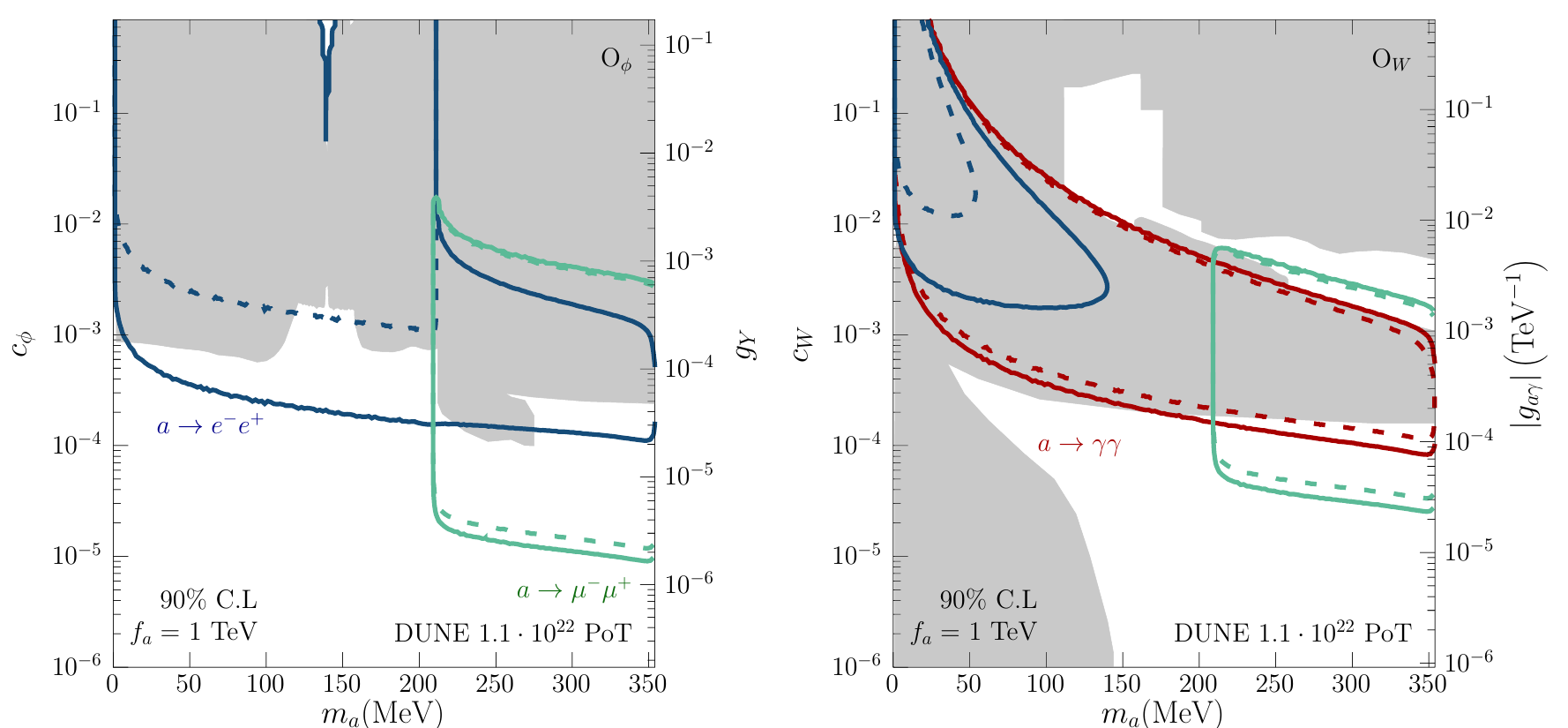}
\end{center}
\caption{\label{fig:dune-sens-EW} DUNE sensitivity projections (one Wilson coefficient switched on at a time) for $c_{\phi}$ (left panel) and $c_{W}$ (right panel) as a function of $m_a$, assuming $f_a=1\;\text{TeV}$. Our sensitivity lines are shown at 90\% CL, individually for each final state topology as indicated by the labels ($a \to \gamma\gamma$, $a\to e^+ e^-$ and $a\to \mu^+ \mu^-$). Solid (dashed) lines correspond to ND-GAr (ND-LAr). Shaded gray regions indicate current bounds, see Sec.~\ref{sec:bounds}. To facilitate the comparison with previous literature we provide on the right axes the corresponding limits to the so-called \emph{photon-dominance} and \emph{fermion-dominance} scenarios~\cite{Agrawal:2021dbo,Beacham:2019nyx} (see Ref.~\cite{Coloma:2022hlv} for the mapping to our Wilson coefficients). 
}
\end{figure}
%%%%%%%%%%%%%%%%%%%%%

Our resulting sensitivity to ALPs coupled through EW operators is shown in Fig.~\ref{fig:dune-sens-EW}. Let us first discuss the results shown in the left panel, obtained for an ALP coupled predominantly through the $\mathcal{O}_\phi$ operator. As outlined in Sec.~\ref{Section:Production}, in this case the ALP tends to decay preferentially to leptons. Once the muon decay channel is open, the ALP lifetime is considerably reduced and the sensitivity to $a\to e^+e^-$ is diminished in the region of large couplings. In the region of small couplings, on the other hand, the dependence on the total lifetime of the ALP approximately cancels and the result depends exclusively on the decay width for a given channel, $a\to e e, \mu\mu$. Overall, we see that DUNE has an excellent opportunity to considerably improve over current constraints for an ALP decaying into $\mu^+ \mu^-$, by more than one order of magnitude, for both ND-LAr and ND-GAr. In the case of $a\to e^+ e^-$, an improvement is only expected for ND-GAr, while the results for ND-LAr are severely affected by the reduced signal efficiency, see Tab.~\ref{tab:effbkg}. Finally, the results for $a\to\gamma\gamma$ are not shown in this case since the reduced branching ratio, combined with the much larger backgrounds expected, yields sensitivities that are not competitive in light of present bounds. 

The right panel in Fig.~\ref{fig:dune-sens-EW} shows the results for an ALP coupled predominantly through the $\mathcal{O}_W$ operator. While in this case the ALP tends to decay predominantly into photons, this channel is affected by much larger backgrounds. Therefore, although the branching ratio into leptons is suppressed for this operator (see Eq.~\eqref{eq:cll}), the final sensitivities obtained for $a\to\ell\ell$ are similar and even surpass those obtained for $a\to\gamma\gamma$ for large masses, when the decay channel $a\to\mu\mu$ is opened. In particular, in the high-mass region we see that DUNE is expected to improve by almost an order of magnitude over current limits by the E137 experiment.

\subsubsection{Charming ALPs}

In this section, we present the sensitivity projections for the charming ALPs model described in Section \ref{sec:charmALPs}, using the DUNE ND with a total exposure of $N_{\text{PoT}}=1.1 \times 10^{22}$. The sensitivity projections are shown in Fig.~\ref{fig:charming_alps_decays} separately for the final states with dominant branching ratios $(a \rightarrow \gamma \gamma, a \rightarrow \pi^{+} \pi^{-} \pi^0)$. The same way as in the previous scenarios considered, our sensitivity contours also exhibit here the typical shape of a visible decay search. For large values of the couplings, the ALPs become very short-lived and decay before reaching the detector, leading to a loss in sensitivity induced by the exponential term in the decay probability. On the other hand, small couplings are suppressed by both the production rate in Eq.~\eqref{eq:decayDalps} and the fact that the ALPs become too long-lived.

For decays into photon pairs, we include both production from kaon decays and from $D$ decays. Conversely, hadronic decay channels are only available for $m_a \gtrsim 3m_\pi$ and therefore only $D\to \pi a$ is considered in this case. In the case of $a\to\gamma\gamma$ it is worth mentioning that in the region below for masses below $m_a < m_K - m_\pi$ the bound for this model is still dominated by the contribution from $K\to\pi a$, since the suppressed production branching ratio for this scenario (see Eq.~\eqref{eq:Kds-charming}) is partly compensated by the large kaon flux available. Again in this case we see that the resulting sensitivities for this decay channel are not competitive with current constraints, with similar results for the two near detectors.

%%%%%%%%%%%%%%%%%%%%%%%%%%%%%%%%%%%%%%%%%%%%
\begin{figure}[ht!]
\begin{center}
  \includegraphics[width=0.8\textwidth]{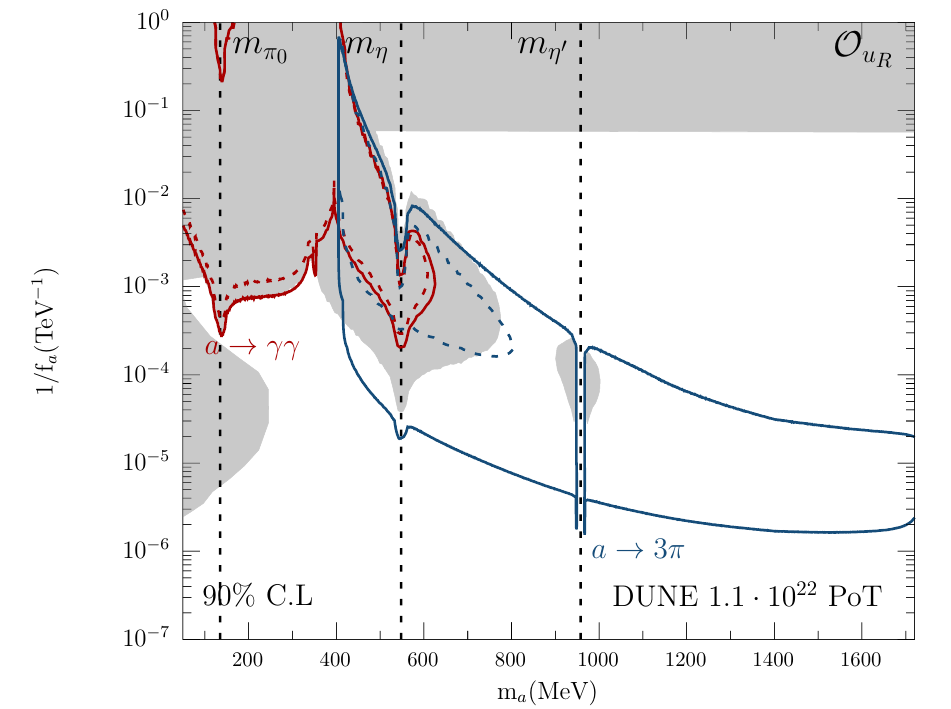}
\end{center}
\caption{
\label{fig:dune-sens-FN} Sensitivity projections for our Benchmark Model III, as described in Sec.~\ref{sec:charmALPs}. Sensitivities are shown separately for a search into di-photon final states and for searches into final states with three pions, which are the dominant decay modes for this scenario. Solid (dashed) lines correspond to ND-GAr (ND-LAr). Shaded gray regions are disfavored by current constraints (see also Fig.~\ref{fig:bounds-FN}). Vertical lines indicate the masses of the neutral pseudoscalar mesons, which explain the sharp features in the sensitivity regions.  
}
\end{figure}
%%%%%%%%%%%%%%%%%%%%%%%%%%%%%%%%%%%%%%%%%%%%

As highlighted in Sec.~\ref{sec:charmALPs} (see Fig.~\ref{fig:charming_alps_decays}) for low masses the branching ratio is dominated by the decay $a\to\gamma\gamma$ whereas hadronic decay channels (driven mainly by $a \to \pi^+ \pi^- \pi^0$) rapidly take over for masses $m_a \simeq 3m_\pi$. Since the $\pi^0$ decays promptly to two photons, the final signal topology for those channels  will be $a \rightarrow \pi^{+} \pi^{-} 2 \gamma$. Other relevant decay modes (not included here for simplicity) would be those involving $\eta$ or $\eta'$ mesons (e.g. $a\to\pi^+ \pi^-\eta$), which also decay promptly into two photons and would lead to a similar signature. From Fig.~\ref{fig:charming_alps_decays} we see that a general improvement is expected over present bounds, specially for ALP masses $m_a > 800~\mathrm{MeV}$. Note that for ALP masses in the vicinity of $m_\eta$ and $m_{\eta^\prime}$ the decay width increases very rapidly, which translates into a sudden loss of sensitivity. 

%%%%%%%%%%%%%%%%%%%%%%%%%%%%%%%%%%%%%%%%%%%%%%%%%%
\section{Summary and conclusions}
\label{sec:conclusions}
The DUNE experiment will be exposed to the LBNF high-intensity neutrino beam. Besides the production of pions, kaons and other light mesons (such as $\eta$ and $\eta'$), the high proton energy at LBNF will gnerate of a significant flux of heavier mesons, such as $D$ and $D_s$. This provides a unique opportunity to study the production of exotic particles feebly coupled to the SM, with masses between $\mathcal{O}(10)~\mathrm{MeV}$ and  $\mathcal{O}(2)~\mathrm{GeV}$. If unstable, these could decay into SM particles and be searched for in the DUNE near detectors, which include both a liquid argon TPC (ND-LAr) and a gaseous argon TPC (ND-GAr).

The potential of DUNE to search for long-lived particles (LLPs) has been studied previously in the literature. However, most of those studies have neglected the impact of backgrounds, arguing that they could be reduced to a negligible level through appropriate selection cuts. In this sense it is worth noting that most studies in the literature consider the ND-GAr detector, which offers a superb environment for these searches thanks to the lower backgrounds expected from neutrino beam interactions. However, ND-GAr is expected to start taking data only during Phase II of the experiment. In light of this, it becomes pressing to assess the capabilities of ND-LAr for this kind of searches, which are a keystone for the BSM program at DUNE.

In this work, we have first computed the background rates according to the expected angular and energy resolution for both ND-GAr and ND-LAr detectors, for decays into pairs of photons, charged leptons or into three pions, with no missing energy. Our results (summarized in Table~\ref{tab:effbkg}) show that these may be reduced to a negligible level for channels involving pairs of muons in the final state; however, the background is significant for LLPs decaying into two photons, which dramatically reduces the expected final sensitivity. Our results also indicate a reduced ND-LAr signal efficiency for decays into $e^+ e^-$ or into three pions. In general, we note that our background study would be applicable to a wide range of BSM models with LLPs, including not only light pseudoscalars (which is the focus of this work), but also light scalars or vector bosons, as considered, for example, in the phenomenological studies in Refs.~\cite{Co:2022bqq, Batell:2023mdn, Brdar:2020dpr, Capozzi:2021nmp, Berryman:2019dme, Dev:2021qjj}.

We have then derived sensitivity regions for generic LLP scenarios using the decay channels outlined above, within a model-independent approach. Our results (shown in Figs.~\ref{fig:bands} and~\ref{fig:br_vs_ctau}, for the ND-GAr) are provided in terms of the production branching ratio and lifetime of the LLP, and may be easily recasted to specific scenarios, including both pseudoscalar and scalar particles. Overall, we find that DUNE has the potential to significantly improve over current constraints for LLP decays into muon pairs, where backgrounds may be reduced to a negligible level while keeping relatively high signal efficiencies. Conversely, we find that searches for decays into two-photons are significantly harder, due to the large backgrounds expected from neutrino interactions, contrary to expectation.

In order to put our sensitivities in context and to compare the potential of DUNE to present bounds, we then study three benchmark models involving axion-like particles (ALPs) coupling to the SM through effective operators: $(i)$ the so-called \emph{gluon dominance} scenario; $(ii)$ a scenario where the ALP is coupled to electroweak operators; and $(iii)$ the so-called \emph{charming ALP} scenario, where the ALP is coupled to right-handed up-quark currents with a non-trivial flavor structure. Our choice of scenarios intends to demonstrate the potential of DUNE to constrain generic ALP models inducing significant couplings not only to photons (as commonly studied in the literature) but to charged leptons and mesons as well, over a wide range of masses. Overall we find that both ND-LAr and ND-GAr will be able to improve significantly over current limits in certain regions of parameter space, for decay channels involving muon pairs or multiple pions. This is particularly so in the case of ALPs produced from $D$ decays, where the parameter space is largely unconstrained by laboratory experiments due to the intrinsic difficulties associated to the production of $D$ mesons. However, we note that the sensitivities in the region where the ALP decays preferably into photon pairs would be affected by the large backgrounds expected, and improving over current limits will be challenging for DUNE. Finally, we stress that if the ALPs decay preferraly into $e^+ e^-$ pairs, ND-GAr will be needed in order to improve over current constraints, while the capabilities of ND-LAr are somewhat limited by the assumed signal efficiency. 

In summary we have shown that, thanks to the \emph{high-intensity}, \emph{high-energy} proton beam available at LBNF, and in combination with the excellent capabilities of the \emph{argon gas TPC near detector}, DUNE has the potential to improve significantly over current constraints for a wide landscape of BSM models with light unstable long-lived particles. We stress that similar searches using the liquid Argon TPC near detector may  be possible. However, the higher backgrounds and lower efficiencies expected translate into significant reductions in sensitivity in most cases. In this regard, the availability of a gas TPC at DUNE will be a key asset in order to ensure the leadership of DUNE on LLP searches at neutrino facilities. 

We emphasize that this work uses only publicly available information from the DUNE collaboration, and that the conclusions are our own and do not represent necessarily the official views of the DUNE Collaboration.

\vspace{2mm} \textbf{Note added:} After this paper was finished, the authors of Ref.~\cite{DallaValleGarcia:2023xhh} pointed out the importance of effects from ALP mixing with neutral pseudoscalar mesons due to the coupling between ALPs and quarks. Although the focus of Ref.~\cite{DallaValleGarcia:2023xhh} is the fermion universal case, such effects would also be relevant for the benchmark models considered here. Mixing effects may lead to an increase in the expected sensitivity for DUNE in certain regions of parameter space, as new production mechanisms become available; however, previous constraints would also have to be reevaluated accordingly. We leave a detailed study of mixing effects between ALPs and neutral pseudoscalars (including a proper reevaluation of past constraints) for future work. 

%%%%%%%%%%%%%%%%%%%%%%%%%%%%%%%%%%%%%%%%%%%%%%%%%%
\begin{acknowledgements}
We warmly thank Pilar Hern\'andez for her involvement in the early stages of this work, and Carlos Pena for illuminating discussions. We also thank Adri\'an Carmona, Felix Kahlhoefer, Luca Merlo, Laura Molina-Bueno, Christiane Scherb and Edoardo Vitagliano for useful discussions and comments. This work has received partial support from the European Union’s Horizon 2020 Research and Innovation Programme under the Marie Sklodowska-Curie grant agreement no.\ 860881-HIDDeN and the Marie Sklodowska-Curie Staff Exchange grant agreement no.\ 101086085–ASYMMETRY. 
PC acknowledges partial financial support by the Spanish Research Agency (Agencia Estatal de Investigaci\'on) through the grant IFT Centro de Excelencia Severo Ochoa no.\ CEX2020-001007-S and by the grant PID2019-108892RB-I00 funded by MCIN/AEI/10.13039/501100011033. She is also supported by grant RYC2018-024240-I, funded by MCIN/AEI/10.13039/501100011033 and by ``ESF Investing in your future''.
JM-A acknowledges support from the Plan GenT programme (grant CIDEGENT/2019/049) funded by the Generalitat Valenciana, and from the Ramón y Cajal programme (grant RYC2021-033265-I) funded by the Spanish MCIN/AEI/10.13039/501100011033 and by the EU (NextGenerationEU/PRTR).
SU acknowledges support from Generalitat Valenciana through the plan GenT programme (CIDEGENT/2018/019) and from the Spanish Ministerio de Ciencia e Innovaci\'on through the project PID2020-113644GB-I00.
\end{acknowledgements}

%%%%%%%%%%%%%%%%%%%%%%%%%%%%%%%%%%%%%%%%%%%%%%%%%%
\appendix
\section{Bounds on {\boldmath $D\to\pi a$} from a reinterpretation of {\boldmath $D\to \tau \nu$} data}
\label{app}
This appendix contains additional details regarding our fit to the BESIII data from  Ref.~\cite{BESIII:2019vhn}. The collaboration divides their data in two samples, which are fitted simultaneously: a $\mu$-like sample, which is dominated by the $D^+\to \mu^+ \nu$ contribution; and a $\pi$-like sample, which contains the $D^+\to \tau^+ \nu, \tau^+ \to \pi^+ \bar\nu$ decays. In particular, their data is binned in $\MM = E^2_\mathrm{miss} - |\vec p_\mathrm{miss}|^2$, where $E_\mathrm{miss} = E_\mathrm{beam} - E_{\mu (\pi)}$. Data is presented in their Fig.~3 using equally sized bins of width $\Delta \MM = 0.02~\mathrm{GeV}^2$, in the range $\MM \in \left[-0.35, 0.35\right]~\mathrm{GeV}^2$. 

The $D^+ \to\tau^+\nu$ measurement at BESIII suffers from five main background components: $D^+ \to K_L^0 \pi^+, D^+\to \pi^0 \pi^+, D^+ \to K_S^0 \pi^+, D^+ \to \eta \pi^+$, and a so-called ``smooth'' background component, that comes mainly from other $D$ decays (such as semileptonic decays) and continuum events. An additional (subleading) contribution to the number of events comes from $D^+ \to \tau^+ \nu$ where the $\tau^+$ does not decay into $\pi^+ \bar\nu$ but to other final states. 

In our fit, we define a $\chi^2$ that fits simultaneously their data in the $\mu$-like and $\pi$-like samples. Specifically we define a binned $\chi^2$ function that depends on a set of nuisance parameters $\xi$:
\begin{equation}
\label{eq:chi2-BESIII}
    \chi^2 (\left\{ \xi \right\}) = \sum_{i,c,s} \frac{\left[\bar n_{i,c,s} - (1 + \xi_c) n_{i,c,s}\right]^2}{\sigma_{i,c,s}^2} \, , 
\end{equation}
where $n_{i,c,s}$ ($\bar n_{i,c,s}$) indicates the predicted (observed) number of events for a given contribution $c$ to a given sample $s$ ($s$=$\mu$-like, or $\pi$-like) in the $i$-th bin. The data points and error bars are taken from Fig.~3 in Ref.~\cite{BESIII:2019vhn}, and the predicted event rates are taken as the best-fit curves provided in the same figure. The final $\chi^2$ is obtained after minimization over the nuisance parameters in Eq.~\eqref{eq:chi2-BESIII}. In doing so, we restrict the $\MM$ range to $\left[ -0.16, 0.19\right]~\mathrm{GeV}^2$ for the $\mu$-like sample, and to $\left[ -0.16, 0.15\right]~\mathrm{GeV}^2$ for the $\pi$-like sample. This leaves us with 17 bins (15 bins) for the $\mu$-like ($\pi$-like) sample. The motivation behind this choice is to use only those points for which the error bars can be accurately extracted from the plot, avoiding the region at high $\MM$ where the error bars are not provided on a linear scale. This is expected to mostly impact the fit to the $K_L^0\pi^+$ background; however we have checked that our choice of energy range allows the fit to constrain its normalization from the information of the $\mu$-like sample.

When fitting the size of the $D^+\to\tau^+\nu$ signal, the collaboration fixes the contributions from $D^+$ decays into $\mu^+\nu$, $\pi^0\pi^+$, $\eta\pi^+$, $K_S^0\pi^+$, while they leave free the normalization of the $K_L^0\pi^+$ and that of the smooth background. Thus, in our fit we proceed in the same way and set
\begin{equation}
   \xi_{D\to\mu\nu} = \xi_{D\to\pi \pi} = \xi_{D\to\eta\pi} = \xi_{D\to K_S\pi} = 0  \quad \mathrm{(fixed)} 
\end{equation}
while we leave completely free the four nuisance parameters associated to $D\to\tau\nu, \tau\to\pi\nu$, $D\to\tau\nu,\tau \to \mathrm{non}-\pi\nu$ , $D \to K_L \pi$, and to the smooth background component. With this procedure, we first check that the resulting $\chi^2$ gives a good fit to the extracted data within the error bars provided in Ref.~\cite{BESIII:2019vhn}, considering the number of degrees of freedom in the fit ($n_\mathrm{dof}$). At the best-fit, we obtain 
\[ \chi^2_\mathrm{min} = 29 \quad \mathrm{for} \quad n_\mathrm{dof} = 28 \, .
\]
This leads to a $p$-value of 41\%, indicating good compatibility with the data. Indeed, our best-fit curves match very well those in Ref.~\cite{BESIII:2019vhn}, as they should. Next, we have checked that we approximately reproduce their fit to the $D^+ \to\tau^+ \nu$. Specifically, we obtain a best-fit for the number of events
\begin{equation}
    N_{ev}(D\to\tau\nu) = 128 \pm 40  
\end{equation}
which corresponds to an overall precision of about $30\%$. This reproduces reasonably well the result of the collaboration ($137\pm 27$ events) albeit with a larger error bar in our case. Thus, our limits may be taken as conservative, keeping in mind that a dedicated analysis (including all data obtained in the full energy range, and a more sophisticated implementation of systematic uncertainties) would probably lead to a better result.

Given the good agreement between our result and that obtained in Ref.~\cite{BESIII:2019vhn} for the SM case, we then proceed to add a contribution from $D\to\pi a$ events as a function of $m_a^2$ (which can be can be directly identified with $\MM$). To this end, we approximate the energy resolution of the detector in $\MM$ by fitting the distribution of $D\to\mu\nu$ to a Gaussian. We believe this should be approximately correct, given that the signal in this case should be a delta function centered at zero. We obtain the best-fit for a Gaussian with mean $\sigma \simeq 0.026~\mathrm{GeV}^2$ and a small bias in its central value, $\mu = 0.016~\mathrm{GeV}^2$ (in what follows, we assume both parameters are independent of $\MM$ in the energy range under consideration). The differential distribution of the expected number of events from $D\to\pi a$ in each sample $s$ is computed for each value of the ALP mass, as:
\begin{equation}
\label{eq:signal-BESIII}
    \frac{dN_{s}(D\to\pi a)}{d\MM} = \epsilon_s \epsilon_\pi N_D \mathrm{BR}_{D\to\pi a}(f_a, m_a^2) \frac{1}{\sqrt{2\pi\sigma^2}} e^{-\frac{(m_a^2 - \MM - \mu)^2}{2\sigma^2} } P_{exit}(f_a, m_a)  \, ,
\end{equation}
where $N_D$ stands for the total number of $D$ mesons produced, while $\epsilon_s$ is the efficiency associated to each sample, $\epsilon_\pi$ is the pion detection efficiency (which we set to $\epsilon_\pi= 90\%$ according to Ref.~\cite{BESIII:2019vhn}), and $P_{exit}$ is the probability for an ALP to exit the BESIII detector before decaying. Regarding the selection efficiencies, we take $\epsilon_{\pi-\mathrm{like}}=44\%$, $\epsilon_{\mu-\mathrm{like}} = 1 - \epsilon_{\pi-\mathrm{like}}$ following Ref.~\cite{BESIII:2019vhn}. The probability $P_{exit}$ is estimated assuming that the parent $D$ mesons are produced in pairs, in collisions with a center-of-mass energy $\sqrt{s}=3.77~\mathrm{GeV}$, and taking the detector radius $\sim 3~\mathrm{m}$, following Ref.~\cite{BESIII:2009fln}. Finally, the number of $D$ mesons is estimated from the observed number of events and the branching ratio reported by the collaboration in Ref.~\cite{BESIII:2019vhn}, taking $N_{ev} = N_D \mathrm{BR}(D \to \tau\nu) \mathrm{BR}(\tau\to\pi)\epsilon_\pi$, $\mathrm{BR}(D\to\tau\nu)_{exp} = 1.2\times 10^{-3}$ and $N_{ev} = 137$.

The contribution of the number of $D\to\pi a$ events to each bin in $\MM$ is obtained integrating Eq.~\eqref{eq:signal-BESIII} within the limits of each bin. These are subsequently added to the total number of predicted signal events, as an additional contribution to the $\chi^2$ in Eq.~\eqref{eq:chi2-BESIII}. A limit can then be obtained on the production branching ratio $\mathrm{BR}(D \to \pi a)$, after marginalization over the same set of nuisance parameters as in the SM case. The resulting limit is shown in Fig.~\ref{fig:bounds-FN} at 90\% CL, for a $\chi^2$ with $n_\mathrm{dof}=28$. 

%%%%%%%%%%%%%%%%%%%%%%%%%%%%%%%%%%%%%%%%%%%%%%%%%%
\begin{figure}[t!b!]
\centering
\includegraphics[width=0.80\textwidth]{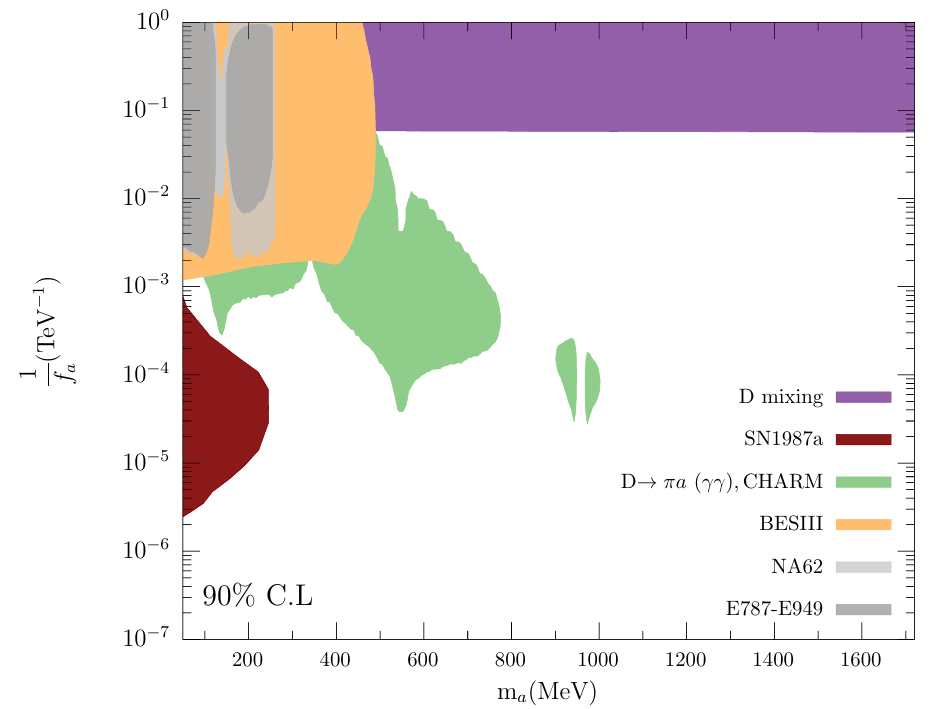}
\caption{\label{fig:bounds-FN} Summary of present bounds on applicable to the Charming ALP scenario, as a function of $m_a$, for the FN-inspired model (see Eq.~\eqref{eq:textureFN}) taking $f_a / \Lambda = \epsilon = m_c/m_t$. The different regions indicate current bounds from NA62~\cite{NA62:2021zjw,NA62:2020xlg}, E787 \& E949~\cite{BNL-E949:2009dza}, CHARM~\cite{BERGSMA1985458}, and BESIII~\cite{BESIII:2019vhn}, all of which have been computed here, see Sec.~\ref{sec:bounds} for details. On the other hand, bounds from SN1987a and $D-\bar D$ mixing are directly taken from Ref.~\cite{Carmona:2021seb}.  
}
\end{figure}
%%%%%%%%%%%%%%%%%%%%%%%%%%%%%%%%%%%%%%%%%%%%%%%%%%

\bibliographystyle{JHEPmod}
\bibliography{references}

\end{document}